\documentclass[5p,times]{elsarticle}

\usepackage{hyperref}

\journal{Parallel Computing Journal (PARCO)}

\usepackage{amssymb}
\usepackage{amsmath}
\usepackage{array}
\usepackage{url}
\usepackage{cleveref}
\usepackage{multirow}
\usepackage{xspace}
\usepackage{todonotes}
\usepackage{color}
\usepackage[procnames]{listings} 
\newcommand{\prettysmall}{\fontsize{6.7}{6.7}\selectfont}
\definecolor{gray98}{rgb}{0.98,0.98,0.98}
\definecolor{gray20}{rgb}{0.20,0.20,0.20}
\definecolor{gray25}{rgb}{0.25,0.25,0.25}
\definecolor{gray16}{rgb}{0.161,0.161,0.161}
\definecolor{gray60}{rgb}{0.6,0.6,0.6}
\definecolor{gray30}{rgb}{0.3,0.3,0.3}
\definecolor{bgray}{RGB}{248, 248, 248}
\definecolor{amgreen}{RGB}{77, 175, 74}
 \definecolor{amblu}{RGB}{72, 88, 102}
\definecolor{amblu}{RGB}{55, 126, 184}
\definecolor{amred}{RGB}{228,26,28}
\newcommand{\lstbg}[3][0pt]{{\fboxsep#1\colorbox{#2}{\strut #3}}}
\lstdefinelanguage{diff}{
  basicstyle=\ttfamily\prettysmall,
  morecomment=[f][\lstbg{red!20}]-,
  morecomment=[f][\lstbg{green!20}]+,
  morecomment=[f][\textit]{@@},
}

\newcommand{\lcudasyntex}{\lstinline[columns=fixed]{<<<}\xspace}
\newcommand{\rcudasyntex}{\lstinline[columns=fixed]{>>>}\xspace}

\usepackage{amssymb}
\usepackage{mathptmx}
\usepackage{graphicx}
\usepackage{amsmath}
\usepackage{url}
\usepackage{xspace}
\usepackage{listings}
\usepackage{subfig}
\usepackage{float}
\usepackage{url}
\usepackage{textcomp}
\usepackage{caption}
\DeclareCaptionType{copyrightbox}
\usepackage{multirow}
\usepackage{pdflscape}
\definecolor{lightgrey}{rgb}{0.905,0.905,0.905}

\newcommand{\gko}{\textsc{Ginkgo}\xspace}

\newcommand{\showchanges}{1}
\ifnum 1=\showchanges
    \definecolor{grn}{rgb}{0, 0.8, 0}
    \usepackage{soul}  
     \newcommand{\del}[1]{\textcolor{red}{}}
    \newcommand{\note}[1]{\textcolor{grn}{#1}}  
\else
    \newcommand{\del}[1]{}
    
    \newcommand{\note}[1]{}
\fi

\begin{document}

\begin{frontmatter}

\title{Ginkgo - A Math Library designed for Platform Portability}

\author{ Terry Cojean$^\star$, Yu-Hsiang Mike Tsai$^\star$, and Hartwig
Anzt$^{\star\dagger}$}
\address{$^{\star}$Steinbuch Centre for Computing, Karlsruhe Institute of
Technology, Germany\\ $^\dagger$Innovative Computing Lab, University of
Tennessee, USA}

\begin{abstract}
The first associations to software sustainability might be the existence of a continuous integration (CI) framework; the existence of a testing framework composed of unit tests, integration tests, and end-to-end tests; and also the existence of software documentation. However, when asking what is a common deathblow for a scientific software product, it is often the lack of platform and performance portability. Against this background, we designed the Ginkgo library with the primary focus on platform portability and the ability to not only port to new hardware architectures, but also achieve good performance. In this paper we present the Ginkgo library design, radically separating algorithms from hardware-specific kernels forming the distinct hardware executors, and report our experience when adding execution backends for NVIDIA, AMD, and Intel GPUs. We also comment on the different levels of performance portability, and the performance we achieved on the distinct hardware backends.
\end{abstract}

\begin{keyword}
Porting to GPU accelerators; Platform Portability; Performance portability; AMD; NVIDIA; Intel
\end{keyword}

\end{frontmatter}


\section{Introduction}
\label{sec:introduction}
When surveying the research software landscape, we can identify some software products that have been successful for several decades~(\cite{dealii,trilinos}). On the other hand, some libraries are successful for a period of time and then fade out. When investigating the source of decline for some products, it is often that the jump from one hardware architecture to the next was too big, and the product failed to keep up with the development of other software and hardware ecosystems.\footnote{Another reason is often the lack of resources for sustained software development, but here we refrain from addressing the topic of under-funding the field of research software engineering.}
In that sense, the lack of software portability and the lack of flexibility to embrace future hardware designs is a time bomb that limits the lifetime of software.

The lack of platform portability becomes even more critical as we see an explosion of diversity in hardware architectures employed in supercomputers.
In the last century, the hardware development was mostly incremental, as it was driven by the clock frequency increase of the processors~(\cite{moore}). During that time, the software developers usually succeeded in transferring to newer chip technologies by applying minor modifications or by simply leveraging the ``free lunch''~(\cite{FreeLunchIsOver}) that came with higher operating frequency. That said, the move from single-core processors to multi-core processors in the early 21st century was incremental enough to be mastered by many software products that did not embrace platform portability as a central design principle. This is partly because using pragmas and the OpenMP language allowed for a smooth transition. In addition, only the performance---not the functionality---of software was endangered when ignoring multi-threaded or multi-core hardware capacity.
In fact, single-threaded software remains functional and can still achieve acceptable performance.
However, at least since the rise of many-core accelerators (e.g., GPUs) and the adoption of special function units and lightweight ARM processors for supercomputing, software libraries can no longer ignore the hardware changes.
As a consequence, the lack of platform portability for emerging and future hardware technology is among the main threats to the sustainability of a given software product.

In the design of the \gko~\cite{ginkgo-joss} open-source software library, we have the burden and privilege to start from scratch, and to apply the lessons learned in other software projects. In this paper we first recall in~\Cref{sec:background} on different levels of portability before we detail in~\Cref{sec:ginkgodesign} how we develop \gko with platform portability as a central design principle. We present in~\Cref{sec:portingamd} and\Cref{sec:portingintel} our strategy for adopting new hardware architectures and report our experiences in porting to new backends, namely AMD GPUs and Intel GPUs. In~\Cref{sec:performance} we present a brief performance evaluation on how architecture-native backends perform in comparison to using platform portability layers. We conclude by summarizing our central findings in~\Cref{sec:conclusion}.

\section{Platform Portability}
\label{sec:background}

\subsection{The Levels of Portability}
There are multiple levels of platform portability. Depending on the use case, platform targets, and objectives, some applications may find it sufficient to restrict themselves to a specific portability level. The first distinct level is \textit{no portability}, where the code compiles and runs for only one type of high-performance computing (HPC) system. The same sort of hardware and compute capabilities are expected. Another option is to support \textit{partial software portability}. An application using such a model will be dependent on some platform model abstraction. For example, the model could expect any CPU type combined with one or more accelerators, either from AMD or NVIDIA. In such a case, a hybrid programming approach featuring a CPU programming model like OpenMP is combined with an accelerator programming model like HIP to ensure portability (and possibly good performance) on the machine. As a more advanced case, one might consider \textit{full software portability}, where the application is able to execute and run on any type of platform, including hypothetical future machines that might feature field-programmable gate arrays (FPGAs). In this case, a practical example is the SYCL programming model, which features compiler backends that support some FPGAs, all mainstream HPC accelerators, and ARM-based hardware. Finally, and especially important for HPC applications, there is the level of \textit{performance portability}, which means that the code will not only compile and run on target platforms, but it will also achieve high efficiency by providing performance close to the machine's total capabilities. To achieve performance portability, one needs good software design practices (e.g., code portability) \textit{and} full command and understanding of the problems inherent in computing unit granularity vs. problem granularity. The latter requires using specific programming techniques to fully express an application's parallelism and scheduling to spread the workload, dynamically, depending on the machine hardware's computing units.

\subsection{Designing for Platform Portability}
Ignoring efforts that are likely doomed to fail, as they permanently redesign to reflect the changes in hardware architecture, one can identify two different approaches to enable cross-platform portability and readiness for future hardware architectures.

\textbf{Relying on a Portability Layer.}
One approach is the adoption of a hardware portability layer that is devoted to supporting hardware through an abstraction. Popular examples are the Kokkos~(\cite{kokkos}) abstraction layer and the Raja~(\cite{raja}) abstraction layer; both are extremely successful in supporting new hardware technologies and providing the users with a unique interface that allows applications to run the same parallel code on very different hardware architectures. Another example is the SYCL programming language which has compilers allowing to target a wide range of hardware.
Relying on a portability layer removes the burden of platform portability from the library developers and allows them to focus exclusively on the development of sophisticated algorithms. This convenience comes at the price of a strong dependency on the portability layer, and moving to another programming model or portability layer is usually extremely difficult or even impossible. Furthermore, relying on a portability layer naturally implies that the performance of algorithms and applications is determined by the quality and hardware-specific optimization of the portability layer. This performance penalty may not always be insignificant, as portability layers usually have a wide user base, and dramatic changes to the interface, logic, or kernel design of the portability layer would likely result in the failure of some applications that rely on the portability layer. Hence, performance portability layers should avoid modifying the design or hardware coverage, which can limit the opportunities to heavily optimize kernels for a new hardware architecture.

\textbf{Natively Supporting Various Hardware Backends.}
The second approach is to decouple the library-core functionality from hardware-specific kernels and support the backends for different hardware (e.g., the \gko software package). From a high-level, one could argue that the second approach takes the first approach and combines the high-level algorithms, the hardware abstraction layer, and the hardware-specific kernel support into a single software product. However, unlike the first approach, this supports hand-written optimized kernels for each hardware architecture. In addition, the abstraction and kernel development being focused on a single product allows for a more consumer-specific kernel design and performance optimization. In consequence, libraries following this path can apply much more aggressive hardware-specific optimization and often achieve higher performance. One reason is that the set of kernels is usually much smaller than what portability layers provide as the hardware-specific backends, because only the kernels required by the library's core algorithms are included. A second reason is that a library has more freedom to phase out support for a specific hardware architecture. This can usually be justified because the dependency on a library is generally much looser than the dependency on a portability layer, and applications ``just'' need to find a new library that provides the same functionality, while the much deeper dependency on a portability layer virtually prohibits moving to an alternative portability layer.
To use this model, a library must be designed with modularity and extensibility in mind. Only a library design that relies on the separation of concerns between the parallel algorithm and the different hardware backends can allow such a feature. The different backends need to be managed and interacted with thanks to a specific interface layer between algorithms and kernels. However, the price for the higher performance potential is high: the library developers have to synchronize several hardware backends, monitor and react to changes in compilers, tools, and build systems, and adopt new hardware backends and programming models. The effort of maintaining multiple hardware backends and keeping them synchronized usually results in a significant workload that can easily exceed the developers' resources.

\begin{figure*}[!t]
  \centering
  \includegraphics[width=0.95\linewidth]{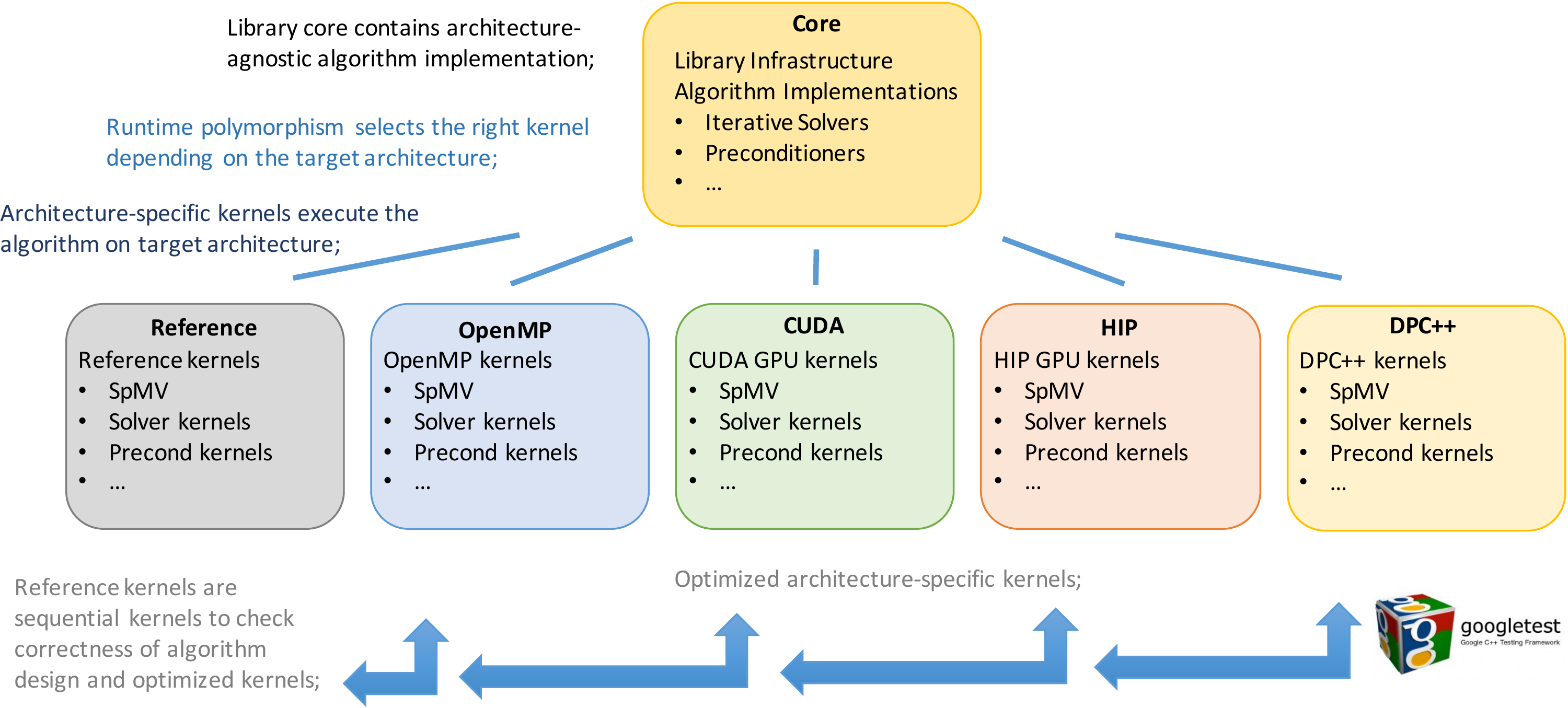}
  \caption{The \gko library design overview.}
  \label{fig:gkodesgin}
\end{figure*}

\section{Developing Gingko using Platform Portability as Central Design Paradigm}
\label{sec:ginkgodesign}
The \gko numerical linear algebra library is developed by acknowledging platform portability as a central design principle. This becomes apparent in the visualization of the software architecture radically separating the library core from the hardware-specific kernels, see \Cref{fig:gkodesgin}. All classes, library logic, and generic algorithm skeletons are accumulated in the library ``core'' which, however, is useless without the driver kernels available in the distinct hardware backends. We note that the ``reference'' backend contains sequential CPU kernels used to validate the correctness of the algorithms and as reference implementation for the unit tests realized using the googletest\cite{googletest} framework. The ``OpenMP'' backend contains kernels for multicore architectures that are parallelized using OpenMP. The ``CUDA'' and ``HIP'' backends are heavily optimized GPU backends written in the hardware-native language: CUDA for NVIDIA GPUs, HIP for AMD GPUs~\cite{isc2019}. Given the significant level of similarity between the ``CUDA'' and the ``HIP'' executor, a shared folder (omitted in the visualization for simplicity) contains kernels that are up to architecture-specific parameter configurations identical for the AMD and the NVIDIA GPUs, respectively~\cite{amdporting}. Collecting those kernels in a folder included by the backends reduces code duplication and maintenance efforts.
The latest addition is the ``DPC++'' backend tailored towards Intel GPUs but usable also on other architectures supporting DPC++ code~\cite{dpcpp}. In opposition to the other backends that are already in production mode, the ``dpc++'' backend is currently under heavy development as also the technical details about Intel's future discrete GPUs are still unknown.
During library configuration, the user decides which backends to compile. The ``executor'' then allows to select the target architecture, the respective kernels are then chosen during execution via dynamic polymorphism.

This is possible as the executor is a central class in \gko{} that provides all important primitives for allocating/deallocating memory on a device, transferring data to other supported devices, and basic intra-device communication (e.g., synchronization).
An executor always has a \texttt{master} executor which is a CPU-side executor capable of allocating/deallocating space in the main memory. This concept is convenient when considering devices such as CUDA or HIP accelerators, which feature their own separate memory space. Although implementing a \gko{} executor that leverages features such as unified virtual memory (UVM) is possible via the interface, in order to attain higher performance we decided to manage all copies by direct calls to the underlying APIs~\cite{ginkgoarxiv}.

\section{Adopting AMD GPUs}
\label{sec:portingamd}
Even though \gko is developed with platform portability as a central design principle, it initially only featured executors for sequential execution (``Reference''), OpenMP-parallelized multicore execution (``OpenMP''), and CUDA-based NVIDIA GPU execution (``CUDA''). Therefore, the addition of a ``HIP'' executor enabling the execution of code on AMD GPUs is the proof-of-concept test for the executor-based platform portability model~\cite{amdporting}.

\textbf{CMake integration.}

When adding an executor based on a new programming language, a first step is the integration of the compiler in the CMake configuration and compilation process.
In the adoption of the HIP ecosystem, we make heavy use of the CMake build system generator. One way of integrating HIP into the build process would be to use the \texttt{hipcc} compiler for the entirety of the project. We chose a less intrusive approach, by relying on the HIP packages which thanks to modern CMake features allows to integrate them via CMake's \texttt{find\_package()} command. AMD provides the following important packages for \gko:
\begin{itemize}
   \item \texttt{hip-config.cmake} etc.,\\ included via \texttt{find\_package(hip)};
   \item \texttt{hipblas-config.cmake},\\  included into a project via \texttt{find\_package(hipblas)} \\ (resp. for \texttt{hipsparse});
   \item \texttt{FindHIP.cmake}, \\included into a project via \texttt{find\_package(HIP)}.
\end{itemize}

The last file provides the main macros for creating HIP projects, and defines a new CMake HIP language to compile the HIP files. The most important macros are \texttt{hip\_add\_executable()} and \texttt{hip\_add\_library()}, similar to the ones declared in\\ \texttt{FindCUDA.cmake} file which provided equivalent CUDA functionality until CUDA became a natively-supported CMake language in CMake version 3.8.

HIP allows to compile code either for CUDA support or for ROCm support, depending on whether a physical GPU is detected or the environment variable \texttt{HIP\_PLATFORM} is set to either \texttt{nvcc} (NVIDIA with CUDA) or \texttt{clang} (AMD with ROCm). During compilation, the HIP header libraries bind to either the CUDA libraries or the ROCm libraries.

Despite providing this convenient setup, currently, several pitfalls exist that require applying customized workarounds. The \texttt{hip-config.cmake} depends on all the ROCm subcomponents, which means that when using the HIP platform \texttt{nvcc}, depending on this package creates too many needless dependencies. In addition, the hipblas package itself always depends on this same file, even on a \texttt{nvcc} platform\footnote{\url{https://github.com/ROCmSoftwarePlatform/hipBLAS/issues/53}}. For convenience, we currently maintain a fork of hipblas\footnote{\url{https://github.com/tcojean/hipBLAS}} that can be accessed by users to shortcut the painful library setup. Some extra asymmetry exists between the two platforms \texttt{clang} (ROCm) and \texttt{nvcc} (CUDA). One such example is the process of locating the CUDA and ROCm libraries. The HIP and ROCm paths are often handled in the main FindHIP.cmake file, whereas finding the CUDA locations in the \texttt{nvcc} case is left to the user.

Another issue is that the \texttt{hip-config.cmake} file hard-codes several AMD device-specific flags which are non-standard, but strictly required when compiling HIP code. These flags get propagated to any dependency (especially when building the \gko framework as a static library), and thus create an error when linking with compilers other than the HIP compiler. Our workaround is to automatically remove these flags when invoking another compiler.

Finally, the \texttt{FindHIP.cmake} struggles with compiling
 shared libraries in complex settings, but throws an exception when using the
 \texttt{AMD}
 backend\footnote{\url{https://github.com/ROCm-Developer-Tools/HIP/issues/1029}}.
 The workaround we apply is to explicitly set the
 \texttt{LINKER\_LANGUAGE} properties of the library to \texttt{HIP}.

We note that AMD and Kitware plan to implement native support for HIP in CMake, such integration would likely fix most of the issues we outlined. Overall, we found that the code skeleton in Listing \ref{code:gkohipcmakelists} is the only one that allows us to successfully integrate HIP code as a subcomponent into a complex project featuring also other subcomponents which rely on CUDA, OpenMP and other libraries.

\begin{figure}[!h]
\begin{center}
\begin{minipage}[t]{0.93\columnwidth}
\begin{lstlisting}[language=bash,caption={Schematic example for
  integrating a HIP module into an existing
  project.},label={code:gkohipcmakelists},morekeywords={hip_add_library,
  STREQUAL, endif, elseif, AND, OR, foreach, endforeach, target_include_directories, REQUIRED,
  target_compile_options, PRIVATE, PROPERTIES, find_package, target_link_libraries,
  target_include_libraries, set, set_target_properties,set_source_file_properties}]
# set ROCM_PATH, HIP_PATH, HIPBLAS_PATH,
# HIPSPARSE_PATH, etc. the path of the
# respective installations if not set,
if(HIP_PLATFORM STREQUAL "nvcc")
  # ensure the CUDA library can be found
  # in the default path or CUDA_PATH is
  # set (requirement of hipcc),
endif()
# set CMAKE_MODULE_PATH and CMAKE_PREFIX_PATH
find_package(HIP REQUIRED)
find_package(hipblas REQUIRED)
find_package(hipsparse REQUIRED)

set(GKO_HIP_SOURCES path/to/source/files.hip.cpp)
set_source_files_properties(${GKO_HIP_SOURCES}
  PROPERTIES HIP_SOURCE_PROPERTY_FORMAT TRUE)
hip_add_library(ginkgo_hip ${GKO_HIP_SOURCES}
  <OPTIONS>)

# On clang backend, set --amdgpu-target
# options when selecting specific GPU targets
if(GKO_HIP_AMDGPU AND HIP_PLATFORM STREQUAL "clang")
  foreach(target ${GKO_HIP_AMDGPU})
    target_compile_options(ginkgo_hip
      PRIVATE --amdgpu-target=${target})
    target_link_libraries(ginkgo_hip
      PRIVATE --amdgpu-target=${target})
  endforeach()
endif()

if(HIP_PLATFORM STREQUAL "clang")
  set_target_properties(ginkgo_hip PROPERTIES
    LINKER_LANGUAGE HIP)
  # Remove device specific flags automatically
  # propagated
  set_target_properties(hip::device PROPERTIES
    INTERFACE_COMPILE_OPTIONS "")
  set_target_properties(hip::device PROPERTIES
    INTERFACE_LINK_LIBRARIES "hip::host")
  target_link_libraries(ginkgo_hip hip::device)
elseif(HIP_PLATFORM STREQUAL "nvcc")
  find_package(CUDA 9.0 REQUIRED)
  target_link_libraries(ginkgo_hip
    ${CUDA_LIBRARIES})
endif()

target_link_libraries(ginkgo_hip roc::hipblas
  roc::hipsparse)
target_include_directories(${HIP_INCLUDE_DIRS}
  ${HIPBLAS_INCLUDE_DIRS}
  ${HIPSPARSE_INCLUDE_DIRS})
\end{lstlisting}
\end{minipage}
\end{center}
\label{fig:gkohipcmakelists}
\end{figure}

\textbf{Porting CUDA code to HIP via the Cuda2Hip script.}
\label{sec:cuda2hip}

For easy conversion of CUDA code to the HIP language, we use a script based on
the hipify-perl script provided by AMD with several modifications to meet our
specific needs. First, the script generates the target filename including the
path in the ``hip'' directory. Then AMD's hipify-perl script is invoked to
translate the CUDA kernels to the HIP language, including the
transformation of NVIDIA's proprietary library functions to AMD's library
functions and the kernels launch syntax. Next, the script changes all
CUDA-related header, namespace, type- and function names to the corresponding
HIP-related names. By default, the script hipify-perl fails
to handle namespace definitions. For example, the hipify-perl script changes\\
\texttt{namespace::kernel\lcudasyntex...\rcudasyntex(...)}\\to\\
\texttt{namespace::hipLaunchKernelGGL(kernel, ...)}\\ while the correct
output would be \\ \texttt{hipLaunchKernelGGL(namespace::kernel, ...)}.
In the Cuda2Hip script, we correct the namespaces generated by the hipify-perl script
after having applied the hipify-perl script to all kernels.\\

\begin{figure}
\centering
    \includegraphics[width=.98\columnwidth]{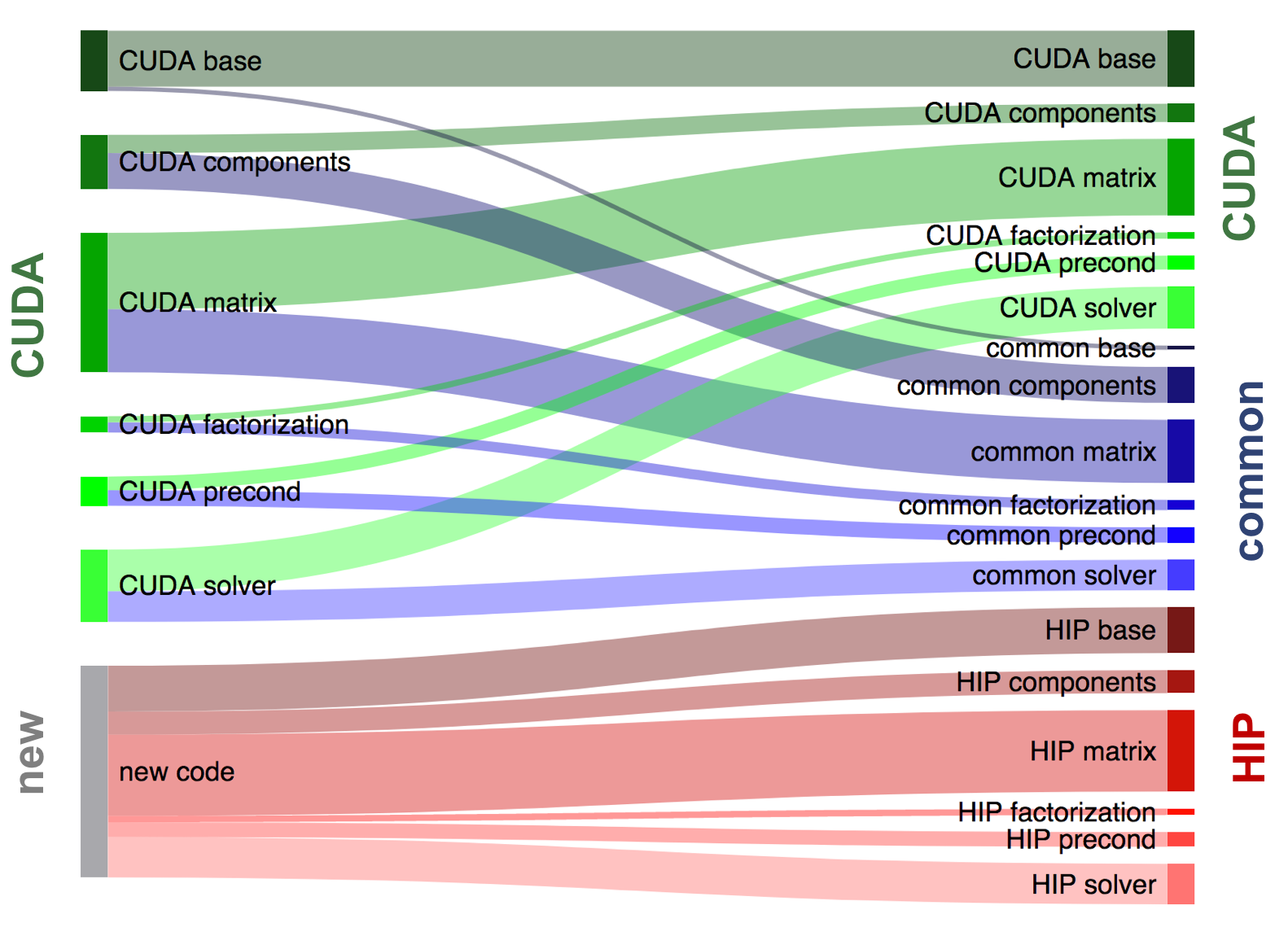}
    \label{fig:reorganization}
    \caption{Reorganization of the \gko library to provide a HIP
      backend for AMD GPUs.}
\end{figure}

\textbf{Avoiding code duplication.}
When adding a HIP backend, we notice a high level of similarity in both kernel design and syntax between the CUDA kernels designed for NVIDIA GPUs and the HIP kernels for AMD GPUs. Thus, the straight-forward addition of a HIP backend would introduce a significant level of
code duplication. For this purpose, we create the ``common'' folder containing
all kernels and device functions that are identical or the CUDA and the HIP
executor except for architecture-specific kernel configuration parameters (such as warp size or
\texttt{launch\_bounds}). These configuration parameters are not set in the
kernel file contained in the ``common'' folder, but in the files located in
``cuda'' and ``hip'' that are interfacing these kernels. The CUDA and HIP backends then include the files in the ``common'' similar to header files and configure the parameters.
Obviously, for separating kernel skeletons shared by the CUDA and HIP executors, the Cuda2Hip script
discussed previously needs to be extended by identifying hardware-specific kernel execution parameters, replacing these with variables, and generating only the kernel call functions in the CUDA and HIP backends, while placing the parameterized kernel skeletons in the common folder that are then included by the executors.
Even though this reorganization requires the derivation of sophisticated scripts,
it pays off as we can avoid
code duplication while still configuring the parameters for optimal kernel
performance on the distinct hardware backends~\cite{amdporting}.
The effect on \gko's code stack is visualized in \Cref{fig:reorganization} where the left-hand side represents the code statistics (lines of code) before the addition of the HIP executor, and the right-hand side the reorganized code with a significant portion of the kernel code ending up in a ``common'' folder shared by the CUDA executor and the HIP executor.\\

\textbf{Cooperative groups.}
CUDA 9 introduced cooperative groups for flexible thread programming.
Cooperative groups provide an interface to handle thread block and warp groups
and apply the shuffle operations that are used heavily in \gko for optimizing
sparse linear algebra kernels. HIP~\cite{hip} only supports block and grid
groups with \texttt{thread\_rank()}, \texttt{size()} and \texttt{sync()}, but no
subwarp-wide group operations like shuffles and vote operations.

For enabling full platform portability, a small codebase, and preserving the
performance of the optimized CUDA kernels, we implement cooperative group
functionality for the HIP ecosystem. Our implementation supports the
calculation of size/rank and shuffle/vote operations
inside subwarp groups. We acknowledge that our cooperative group
implementation may not support all features of CUDA's cooperative group
concept, but all functionality we use in \gko.

\begin{figure*}[!h]
  \centering
  \begin{tabular}{lcr}
    \includegraphics[width=0.55\linewidth]{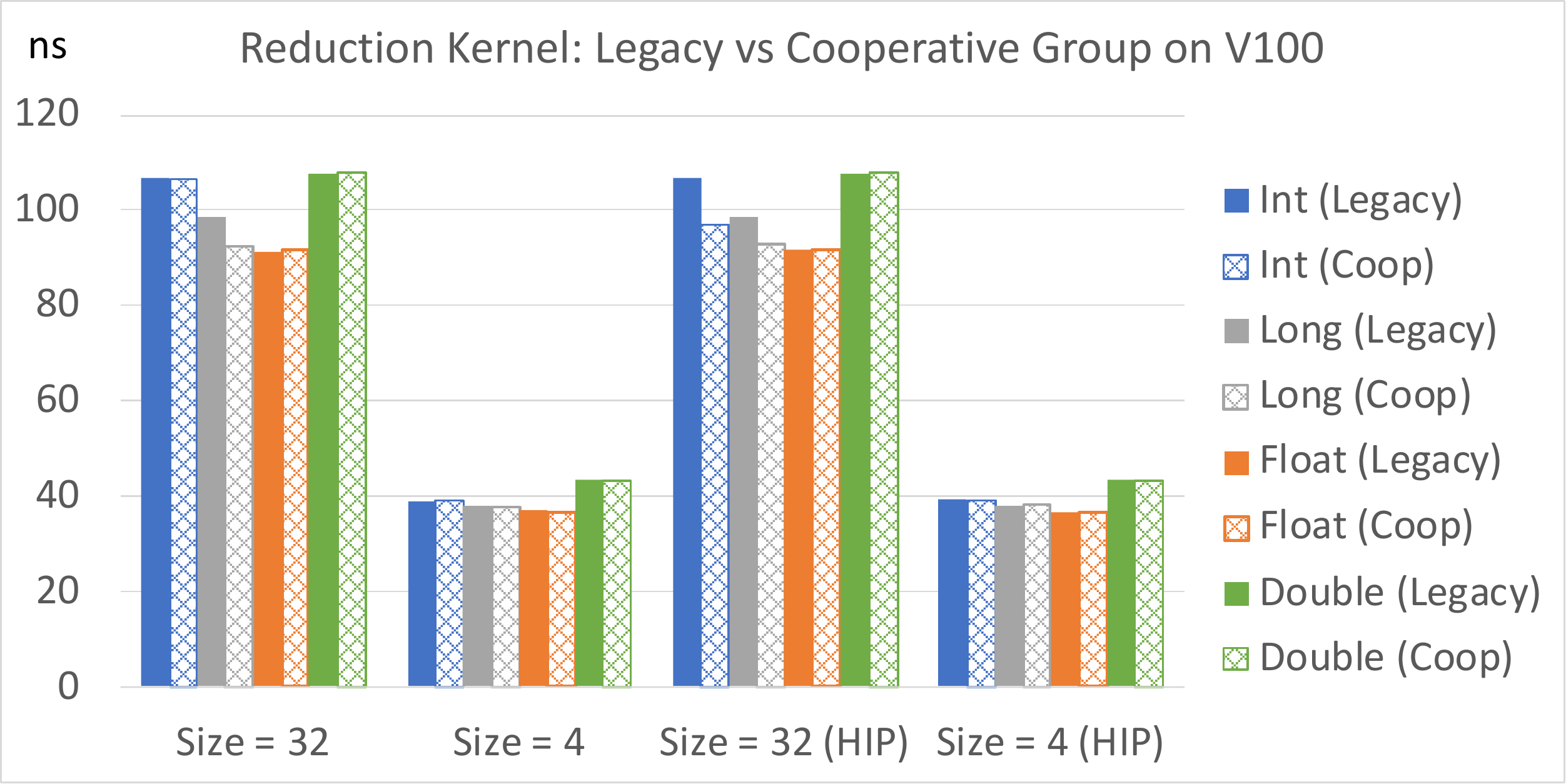}
    & \quad\ \quad &
    \includegraphics[width=0.36\linewidth]{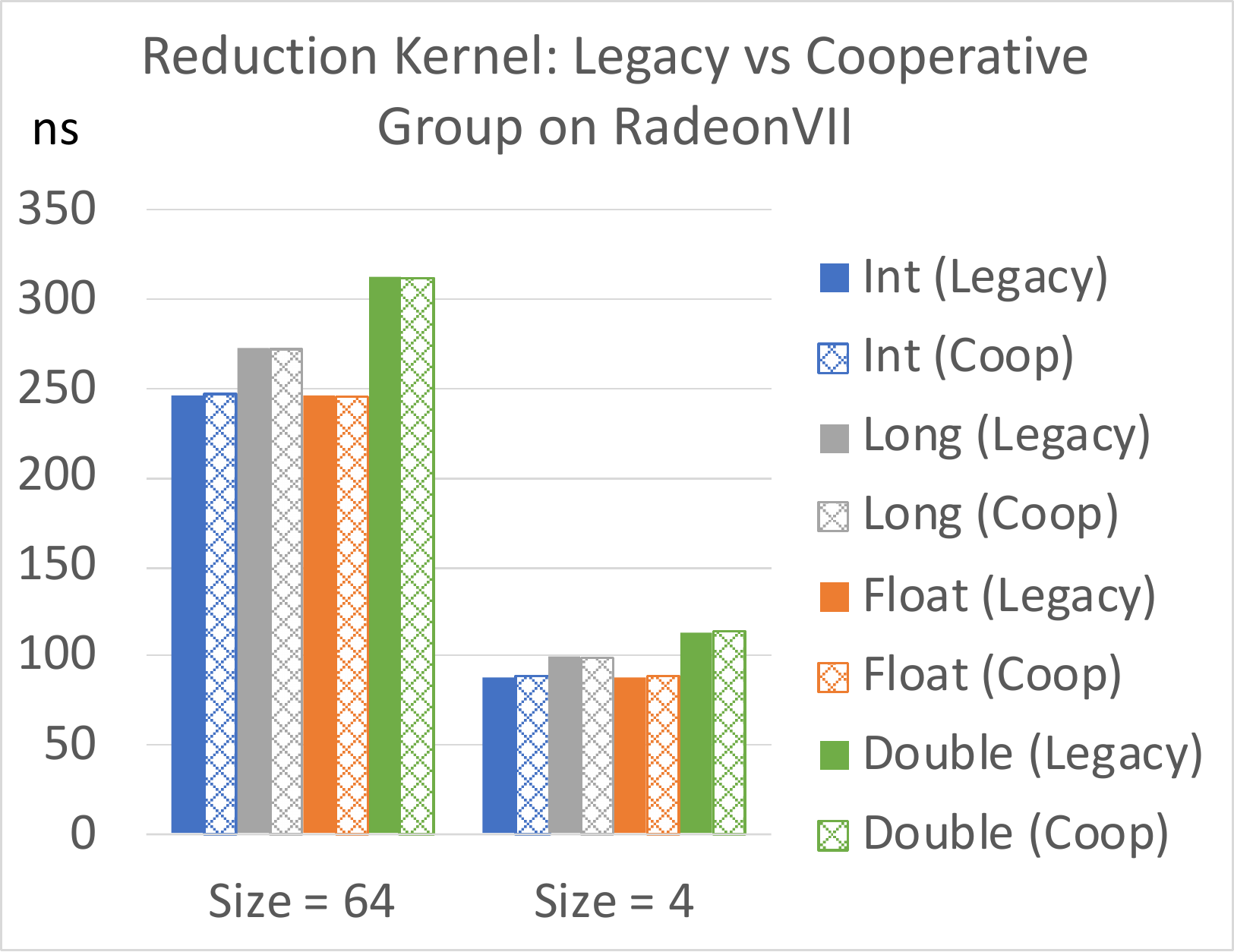}
  \end{tabular}
  \caption{\gko's cooperative groups vs. legacy functions for different data
  types on V100 (left) and RadeonVII (right)~\cite{amdporting}.}
  \label{fig:cooperativeperformance}
\end{figure*}

The cross-platform cooperative group functionality we implement with shuffle
and vote operations covers CUDA's native implementation. HIP only interfaces
CUDA's warp operation without \texttt{\_sync} suffix (which refers to
deprecated functions), so we use CUDA's native warp operations to avoid
compiler warning and complications on NVIDIA GPUs with compute capability 7.x
or higher.
We always use subwarps with contiguous threads, so we can use the block index to
identify the threads' subwarp id and its index inside the subwarp.
We define
\begin{scriptsize}
\[
  \begin{split}
    \texttt{Size} &= \text{Given subwarp size}\\
    \texttt{Rank} &= \texttt{tid \% Size}\\
    \texttt{LaneOffset} &= \lfloor{} \texttt{tid \% warpsize / Size}
\rfloor{}\times \texttt{Size}\\
    \texttt{Mask} &= \texttt{$\sim 0$ >> (warpsize - Size) <<
    LaneOffset}
  \end{split}
\]
\end{scriptsize}

where \texttt{tid} is local thread id in a thread block such that \texttt{Rank}
gives the local id of this subwarp, and \texttt{$\sim 0$} is a bitmask of 32/64
bits, same bits as \texttt{lane\_mask\_type}, filled with 1 bits according to
CUDA/AMD architectures, respectively. Using this definition, we can realize the
cooperative group interface, for example for the \texttt{shfl\_xor}, \texttt{ballot},
\texttt{any}, and \texttt{all} functionality:
\begin{scriptsize}
\[
  \begin{split}
    \texttt{subwarp.shfl\_xor(data, bitmask)} &= \texttt{\_\_shfl\_xor(data,
    bitmask, Size)}\\
    \texttt{subwarp.ballot(pred)} &= \texttt{(\_\_ballot(pred) \&
    Mask) >> LaneOffset} \\
    \texttt{subwarp.any(pred)} &= \texttt{(\_\_ballot(pred) \& Mask)
    != 0 }\\
    \texttt{subwarp.all(pred)} &= \texttt{(\_\_ballot(pred) \& Mask)
    == Mask}
  \end{split}
\]
\end{scriptsize}
Note that we use the \texttt{ballot} operation to implement \texttt{any} and
\texttt{all} operations. The original warp \texttt{ballot} returns the answer for the
entire warp, so we need to shift and mask the bits to access the subwarp
results. The \texttt{ballot} operation is often used in conjunction with bit
operations like the population count (\emph{popcount}), which are provided by
C-style type-annotated intrinsics \texttt{\_\_popc[ll]} in CUDA and HIP. To
avoid any issues with the 64bit-wide lane masks on AMD GPUs, we provide a
single function \texttt{popcnt} with overloads for 32 and 64 bit integers as
well as an architecture-agnostic \texttt{lane\_mask\_type} that provides the
correct (unsigned) integer type to represent a (sub)warp lane mask.
Experimental results have shown that \gko's custom platform portable
cooperative group implementation is highly competitive to the vendor-provided functionality, see \Cref{fig:cooperativeperformance} \cite{amdporting}.

\section{Adopting Intel GPUs}
\label{sec:portingintel}
Intel is the vendor who is expected to provide the exascale Aurora machine, hosted at Argonne National Laboratory which will feature Intel discrete GPUs. To program these new architectures, Intel is focusing on the SYCL programming framework~\cite{sycl}, which was extended with several programming features allowing both better kernel performance and usability. This is packaged into the DPC++ compiler~\cite{dpcpp}, which is part of the Intel oneAPI framework\footnote{\url{https://www.oneapi.com/}}. \gko started the portability effort to the new Intel oneAPI platform in order to target future Intel hardware.

\textbf{CMake integration.}
At the time of writing, the only way the authors found to integrate DPC++ into our framework is to use the DPC++ compiler for the whole compilation process of our library (e.g., by passing \texttt{CMAKE\_CXX\_COMPILER=dpcpp} to CMake). This usage model means that the CMake integration effort is minimal, but on the other hand, there is no proper CMake isolation of the DPC++ submodule, which prohibits to compile any other \gko{} submodule at the same time, except for \texttt{Reference}.\\

\textbf{Adding a DPC++ Executor to \gko{}.}
LikE SYCL, DPC++ is a C++ single-source heterogeneous programming framework for acceleration offload version targeting the full range of acceleration APIs, such as OpenCL. A core concept of these frameworks is the use of command queues which are objects able to schedule kernels on devices with specific execution contexts. By default, SYCL and DPC++ execute kernels asynchronously, and only the destruction of the queue object synchronizes with the device. Memory copies are all implicit thanks to SYCL providing a specific buffer type for registering data, together with the access mode specification of these data inside kernels. Another key aspect of DPC++ is that (like SYCL) it aims at being compilable on most existing devices, ranging from NVIDIA to AMD GPUs,
Intel FPGAs, GPUs, and general purpose processors.

This usage model of standard SYCL/DPC++ does not fit well with the \gko framework, as \gko handles devices via the \texttt{Executor} model (which can be either a CPU or an accelerator). On the other hand, a new DPC++ executor should be able to target any device type. For this purpose, we have decided to add an optional device selection to our DPC++ executor constructor which allows to specify whether the user wants to target a GPU or a CPU (or any other device type). In addition, we decided that a DPC++ executor always comes with a CPU executor to manage the host-side data (as it is needed for GPUs). We acknowledge that this can result in redundant data in case the DPC++ executor targets a CPU device.

Another aspect important to consider when creating a DPC++ executor for \gko is the ``subgroup'' concept DPC++ introduces to represent a SIMD lane on CPUs or a warp/wavefront on GPUs. Depending on the architecture, the \texttt{subgroup\_size} can differ dramatically, from a size of 1 on some simple devices to 4 or 8 for powerful CPUs, to 16, 32, or 64 on some GPUs. While this concept is not completely new -- \gko already uses warp size 32 for NVIDIA GPUs and wavefront sizes 64 for AMD GPUs -- it can heavily affect performance or even cause errors if selecting the \texttt{subgroup\_size} inappropriate for a given hardware architecture. To resolve this, in \gko, we precompile kernels for all possible \texttt{subgroup\_sizes} (from 1 to 64, in powers of two increments), and dynamically select at runtime the kernel version which fits best to the selected DPC++ device.

Since \gko executors need to provide synchronization features, we have decided that one executor is represented by one DPC++/SYCL queue object which exists during the whole lifetime of the executor. This object is automatically managed thanks to a \texttt{unique\_ptr}. The DPC++ queue primitive\\ \texttt{wait\_and\_throw()} can be used for explicit synchronization. A specific aspect of a DPC++ queue is that, by default, no execution order is guaranteed for the objects submitted to it, since it operates asynchronously. This means that when executing three tasks on the same queue, 1) copy: host $\rightarrow$ GPU, 2) a GPU kernel, 3) copy: GPU $\rightarrow$ host, there is no guarantee that the three operations will be executed in this order. To ensure the right execution order, DPC++ introduces the queue property \texttt{in\_order} which can be used. Another solution to this problem is given by manually synchronizing after every DPC++ queue operation.

Finally, to allow explicit memory management, we rely on the new DPC++ concept of Unified Shared Memory, which can be used via the functions for the explicit memory (de)allocation (\texttt{sycl::malloc\_device()}, \texttt{sycl::free()}). When using memory allocated via these functions, the memory copies can be controlled manually via the DPC++ queue itself using the function \texttt{queue.memcpy()}. The limitation of this concept is that it can not handle settings where another non-DPC++ \gko executor (or device) is used in an application (for example, a HIP enhanced device). In this case, DPC++ can not directly use the \texttt{queue.memcpy()} function to copy data to the other executor. To resolve the challenge of copying data from the DPC++ executor to another (non-CPU) executor, we use the workaround of creating a temporary copy on the master executors running on the CPU and controlling the execution of the DPC++ executor and the other device executor, respectively.

\textbf{Kernel programming}

\begin{figure}[!h]
\begin{center}
\begin{minipage}[t]{0.9\columnwidth}
\begin{lstlisting}[language=C,caption={Minimal example of a DPC++ kernel call featuring a subgroup based reduction.},label={code:dpcppsubgroupreduce},morekeywords={submit, memcpy, malloc_device}]
#include <CL/sycl.hpp>

template<int subgroup_size>
[[ intel::reqd_sub_group_size(subgroup_size) ]]
void reduce(float *a, sycl::nd_item<3> i) {
  auto subgroup = i.get_sub_group();
  auto local_data = a[i.get_local_id(2)];
  #pragma unroll
  for (int bitmask = 1; bitmask < subgroup_size;
     bitmask <<= 1) {
      const auto remote_data =
        subgroup.shuffle_xor(local_data, bitmask);
      local_data = local_data + remote_data;
    }
    a[i.get_local_id(2)] = local_data;
}

int main() {
  sycl::queue q{}; // get default device's queue
  float *dev_A;
  float A[32]; // missing: populate data
  dev_A = sycl::malloc_device<float>(32, q);
  q.memcpy(dev_A, A, 32 * sizeof(float)).wait();
  q.submit([&](sycl::handler &cgh) {
    cgh.parallel_for(
      sycl::nd_range<3>(sycl::range<3>(1, 1, 32),
                       sycl::range<3>(1, 1, 32)),
                       [=](sycl::nd_item<3> i) {
                         reduce<8>(dev_A, i);
                       });
    }).wait();
  q.memcpy(A, dev_A, 32 * sizeof(float)).wait();
}
\end{lstlisting}
\end{minipage}
\end{center}
\label{fig:dpcppsubgroupreduce}
\end{figure}

Listing~\ref{code:dpcppsubgroupreduce} showcases a minimal DPC++ example for setting up and running a subgroup based reduction. We only omit in this code the generation of the data the kernel operates on. The code is composed of two parts, lines~3-16 are device-side functions which handle the reduction on a subgroup. The template parameter reflects the subgroup size (line~3) to allow the compiler to unroll the loop in line~9. The second part in lines~18-33 is the main function containing data management as well as the kernel invocation. In some detail, we first use a default queue, which will access the default device (line~19) and can be used for data management as well as submitting tasks. Lines~22-23 reflect the DPC++ Unified Shared Memory usage, where initially the device-side memory is allocated before copying this data from the host to the device. Line~32 similarly copies back the data to the host before kernel completion. Lines~24-31 show the creation and submission of a task that operates in ``parallel-for'' fashion (line~25). The task initially selects its \texttt{nd\_range} parameter with both the global (line~26) and local, or work group level, (line~27) dimensions. Since the global and work group dimensions are identical, this implies that only one work group is instantiated which then processes the whole kernel range. The kernel executed by this ``parallel-for'' is the lambda function defined in lines~28-30 which only calls the reduction kernel on the device data at each respective ``item'' of the parallel for. Note that in this program, we always call the function \texttt{wait()} of the queue to ensure proper ordering of the operations (we could also use the DPC++ property ``in order'' of the queue). Lines~3-16 present the reduction kernel. On line~4, the keyword \texttt{intel::reqd\_sub\_group\_size} is used to notify that this function targets a subgroup with a size depending on the template parameter (in this example the subgroup size is $8$, see line~29). In summary, the code will create one work group containing four subgroups each composed of 8 elements. Line~6 then accesses this subgroup and stores the representation object. All subgroups can then execute a standard reduction in lines~9-15, which is based on \texttt{subgroup.shuffle\_xor} (line~12).

\textbf{Custom porting script based on {\sc dpct}}

\begin{figure*}[!h]
    \begin{center}
    \includegraphics[width=.7\linewidth]{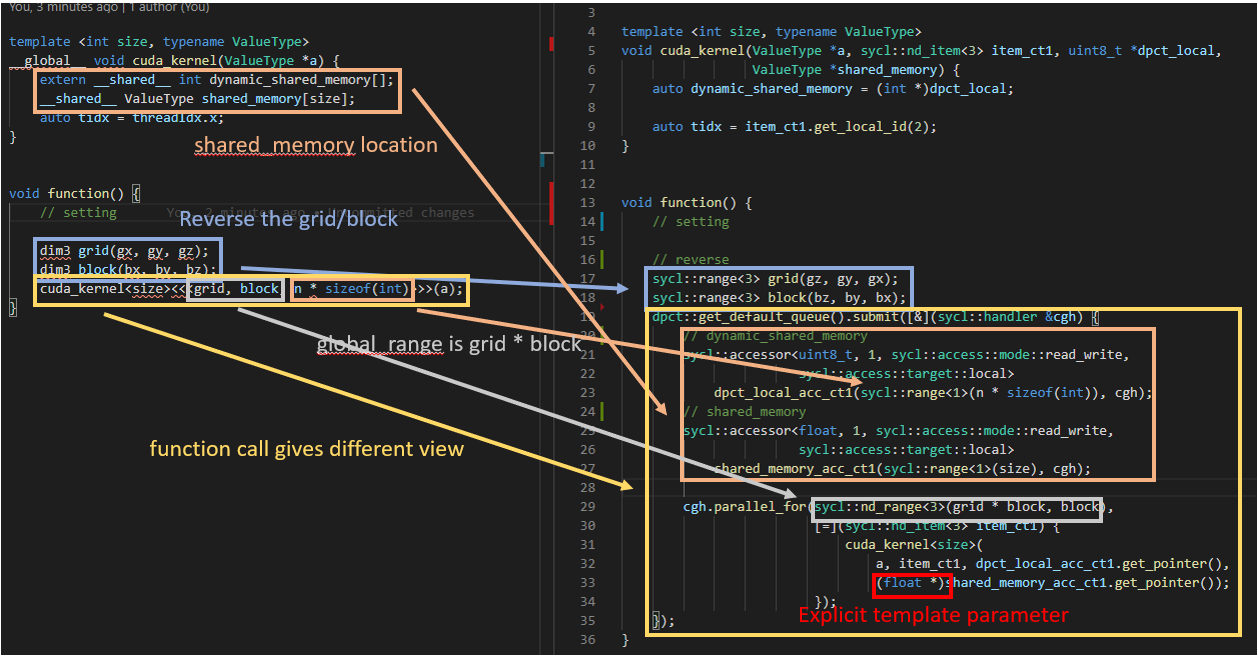}
    \caption{Example of a CUDA code (left) and the {\sc DPCT}-converted code (right) which uses static and dynamic shared memory.}
    \label{fig:dpctconversion}
    \end{center}
\end{figure*}

\begin{figure*}[!h]
   \begin{center}
    \includegraphics[width=.7\linewidth]{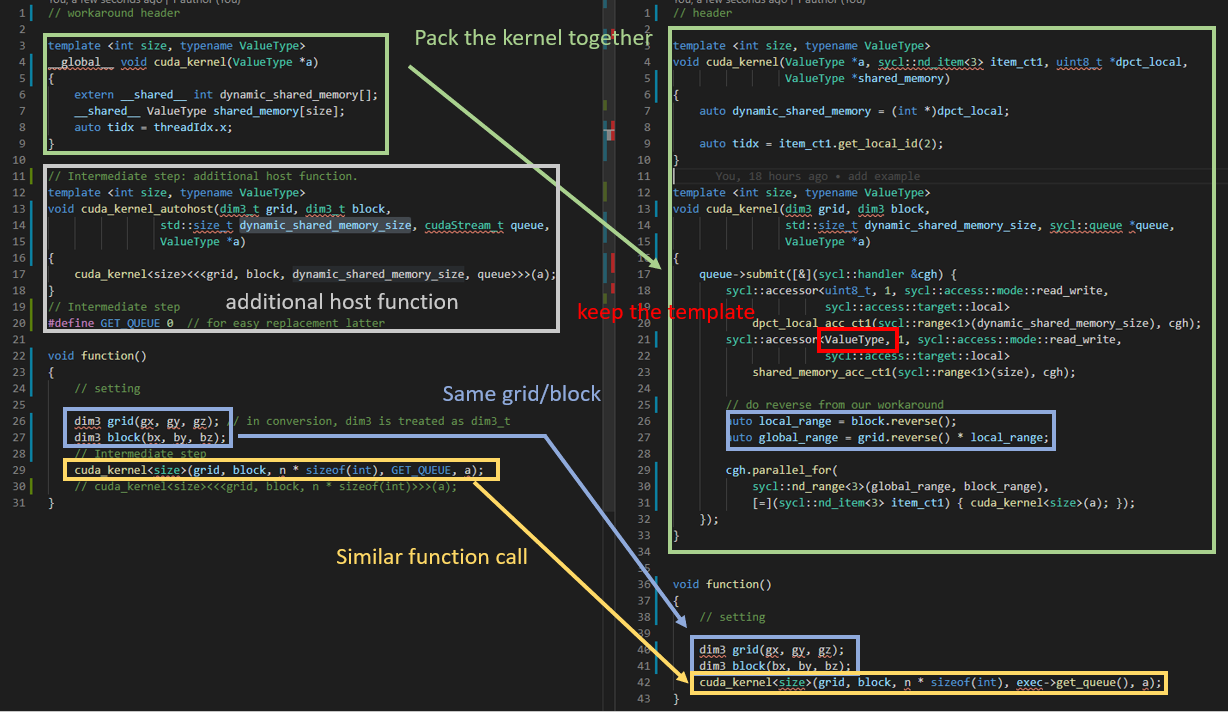}
    \caption{Conversion of the same CUDA code (left) like in Figure~\ref{fig:dpctconversion} to DPC++ (right) via the improved {\sc DPCT} containing our modifications. The CUDA code is also adapted by the script (before conversion only) with the addition of a new host function to force {\sc DPCT} to create a more consistent output.}
    \label{fig:dpctconversion_improved}
    \end{center}
\end{figure*}

Similar to AMD, Intel provides a tool that converts CUDA code to the DPC++ language. This ``DPC++ Compatibility Tool'' ({\sc DPCT}\footnote{\url{https://software.intel.com/content/www/us/en/develop/documentation/intel-dpcpp-compatibility-tool-user-guide/top.html}}) intends to make porting existing CUDA kernels to the OneAPI ecosystem a simple and convenient step. However, at the time of writing, the OneAPI ecosystem is still in its early stages, and the conversion tool has several flaws that require manual fixes. In addition, the tool in some cases requires manual involvement of the programmer as the tool cannot be made general to work for all cases (in these cases a warning is printed by {\sc DPCT}). In the context of developing a DPC++ backend for \gko{}, we enhanced {\sc DPCT} with several additions and error fixes -- some of them customized to the \gko library design, some of them useful for any code conversion. Even though {\sc DPCT} is at the time of writing still under development, we want to list some of these changes and modifications we introduced to customize this tool to our purpose.\footnote{Our porting script can be accessed online \url{https://github.com/yhmtsai/try_oneapi/blob/master/workflow/convert_source.sh}.} We show a concrete example of a standard conversion of a CUDA code using dynamic and shared memory in converted thanks to {\sc DPCT} in Figure~\ref{fig:dpctconversion}, and an intermediate version which is improved thanks to some techniques used by our script in Figure~\ref{fig:dpctconversion_improved}.

We observe in Figure~\ref{fig:dpctconversion} that, by default, DPC++ code has a very different structure and aspect than the CUDA equivalent, and also creates some subtle errors as well. One error concerns the static shared memory type which was based on a template parameter in the left (CUDA) code, but becomes evaluated to the underlying data type (float) in this example on the right (DPC++) code (lines~29 and 33). In addition, there are inconsistencies compared to the CUDA code that are confusing at a first glance. One such issue concerns the grid and block representation which is used in DPC++ (lines~17-18): the dimension which always moves in DPC++ is the right-most one, or ``z'' index, whereas it is the left-most one in CUDA, or ``x'' index. Similarly, DPC++ uses a different concept of global and local range (line~29), where the global range is grid$\times$block. In addition, the DPC++ kernel code is partly embedded in the main host function (the lambda function in the right part, lines~30-34), whereas there is a strict separation in CUDA.

The modifications we apply to {\sc DPCT} as a workaround for these issues are shown in Figure~\ref{fig:dpctconversion_improved}. The main difference is that instead of calling {\sc DPCT} on the original CUDA file, we generate a temporary CUDA file that uses a properly templated extra host function to invoke the kernel itself (lines~11-18 on the left side). In addition, we also create a placeholder \texttt{GET\_QUEUE} (line~20, left) which we pass as \texttt{cudaStream\_t} in the new host function (line~29, left). These two changes allow {\sc DPCT} to generate better and more structured DPC++ code on the right side, where the main host function (lines-36-43, right) looks indeed similar to the original CUDA code. The \texttt{GET\_QUEUE} placeholder can be replaced with our executor's queue accessor by our script (line~42, right), and {\sc DPCT} properly converts the new \texttt{cudaStream\_t} parameter in the new host function to a \texttt{sycl::queue} object. Note that we now need to reverse the \texttt{dim3} grid and block descriptions before using them (lines 26-27, right), which we realize via our custom \texttt{dim3} interface.

\begin{figure*}
  \centering
  \begin{tabular}{lr}
    \includegraphics[width=.92\columnwidth]{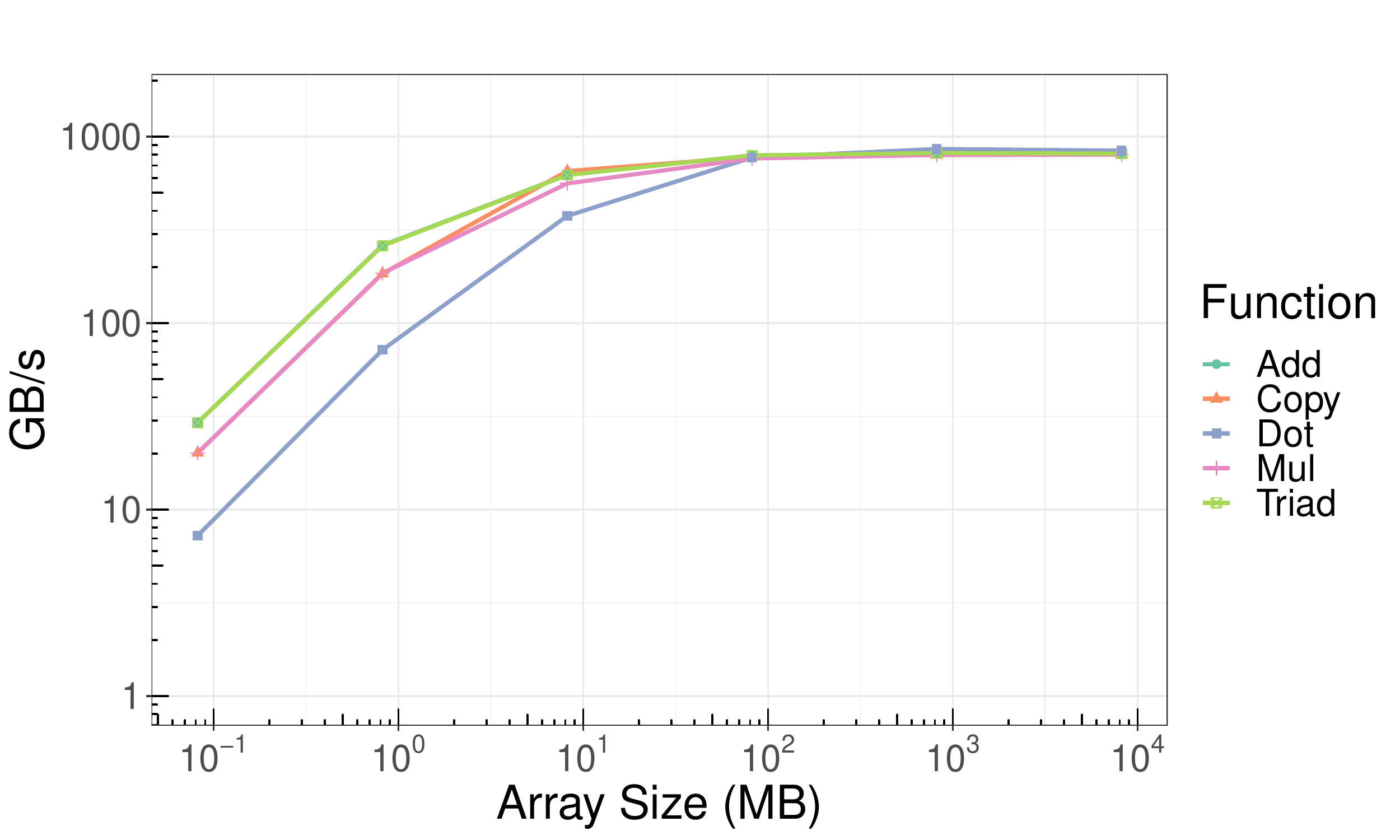}
    &
    \includegraphics[width=.92\columnwidth]{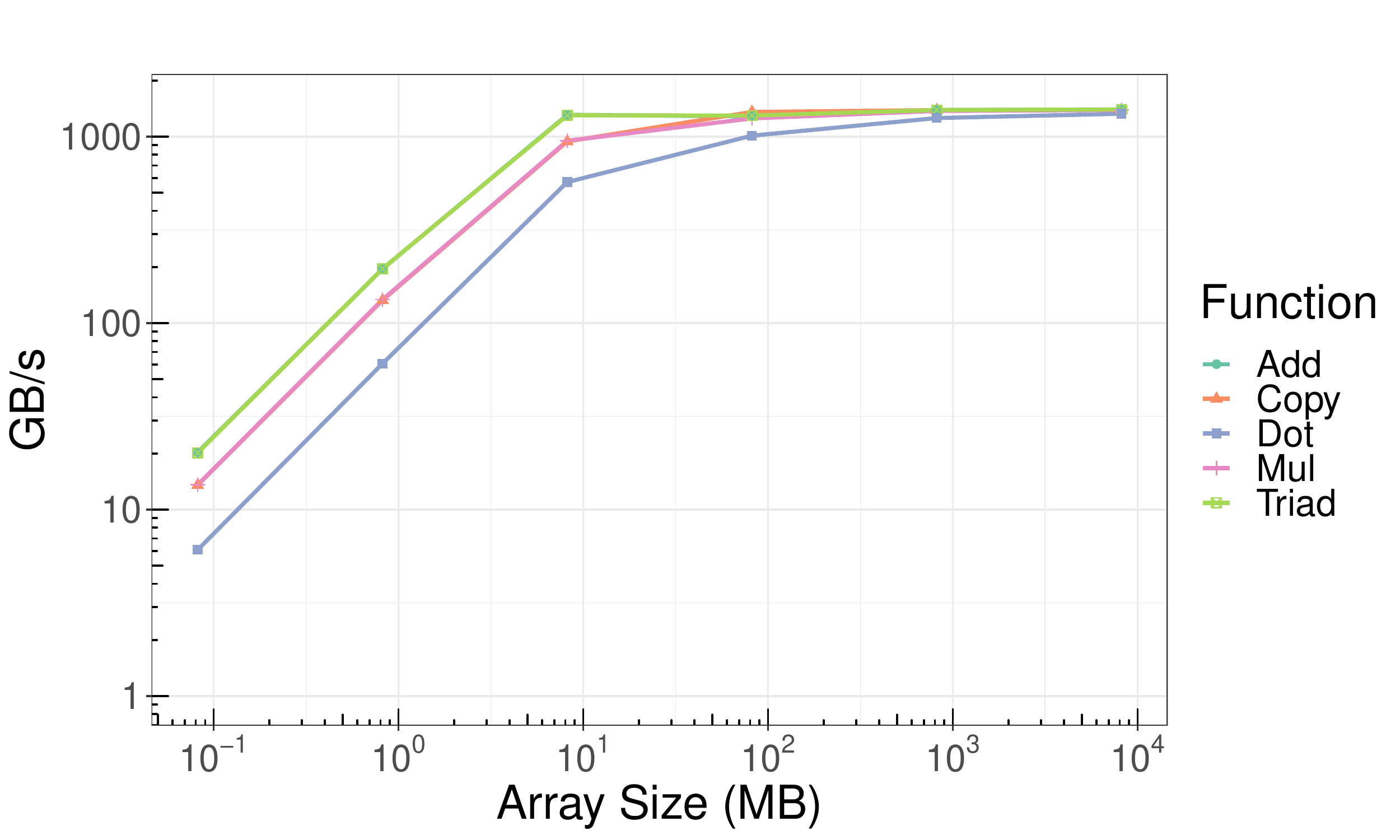}\\
    \includegraphics[width=.92\columnwidth]{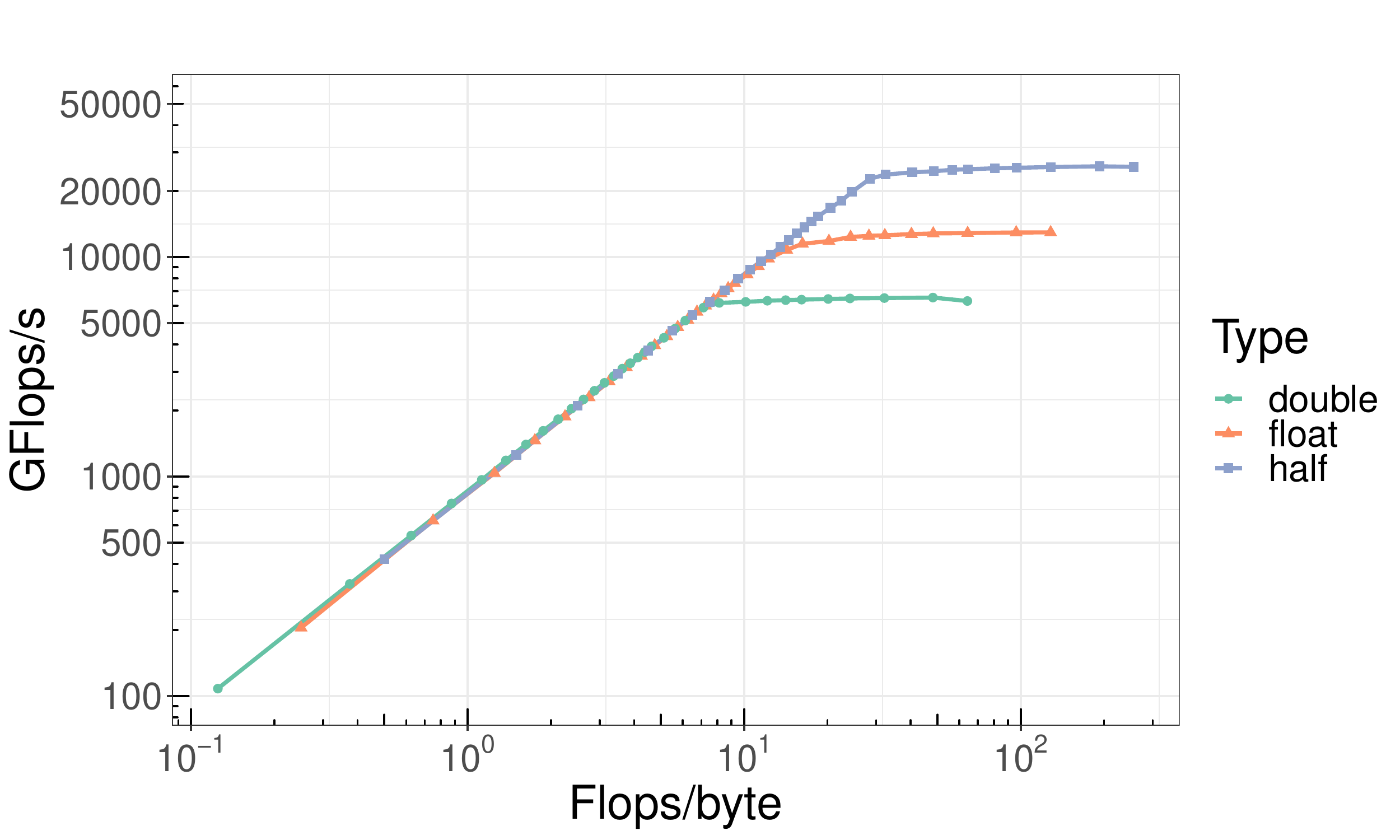}
    &
    \includegraphics[width=.92\columnwidth]{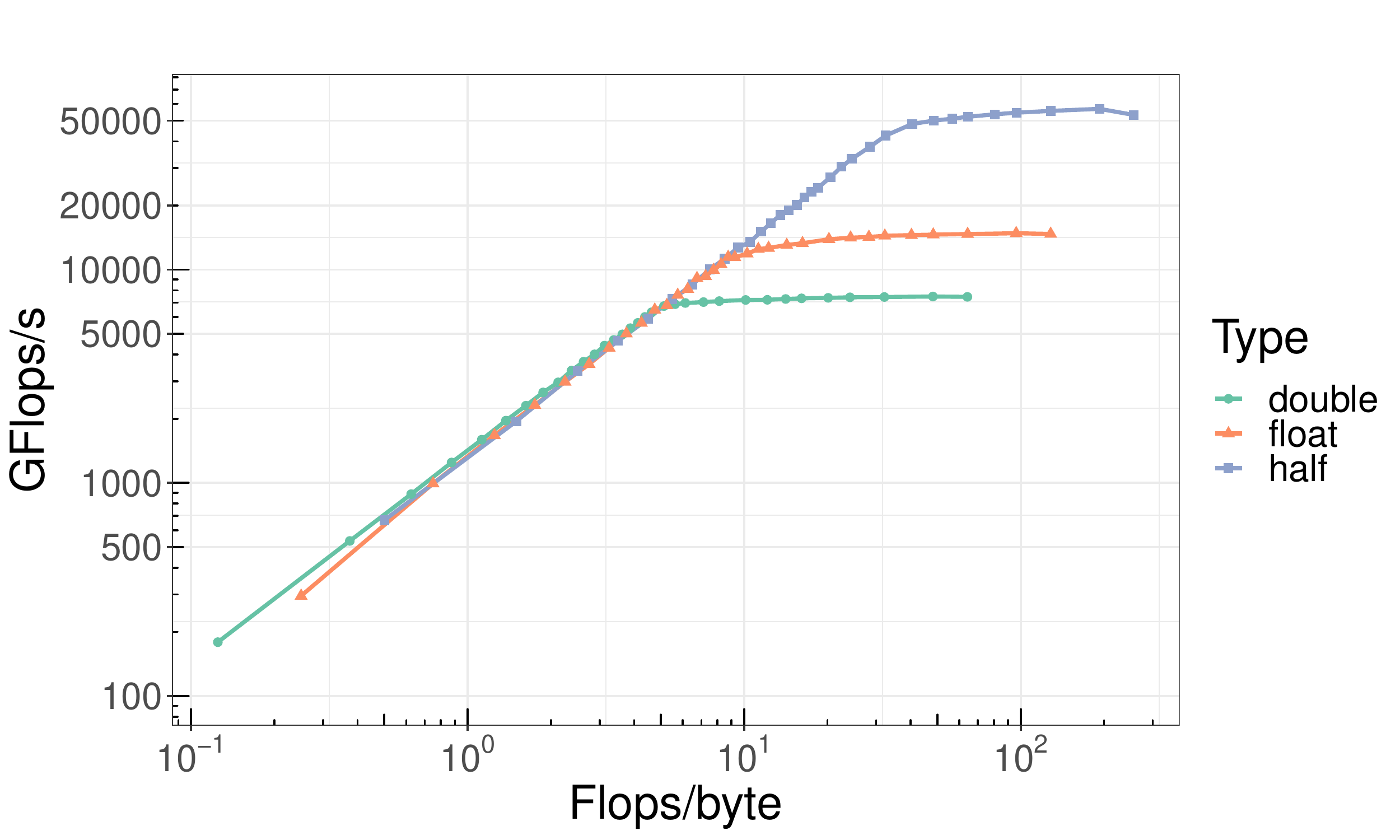}
  \end{tabular}
  \caption{Performance evaluation of the NVIDIA V100 GPU (left) and the NVIDIA A100 GPU (right) using the BabelStream benchmark (top) and the mixbench benchmark (bottom).}
  \label{fig:nvidiaperf}
\end{figure*}

\begin{figure*}
  \centering
  \begin{tabular}{lr}
    \includegraphics[width=.92\columnwidth]{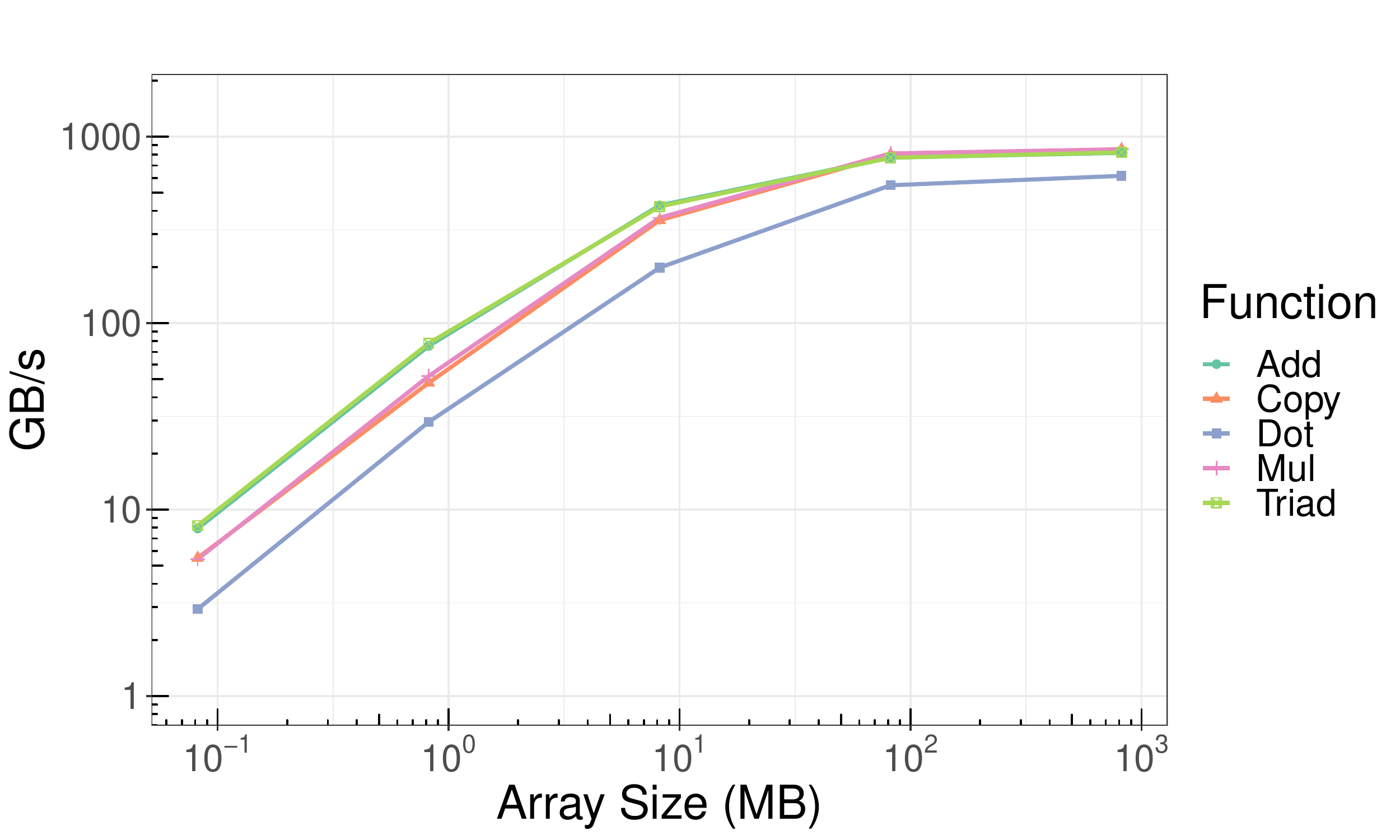}
    &
    \includegraphics[width=.92\columnwidth]{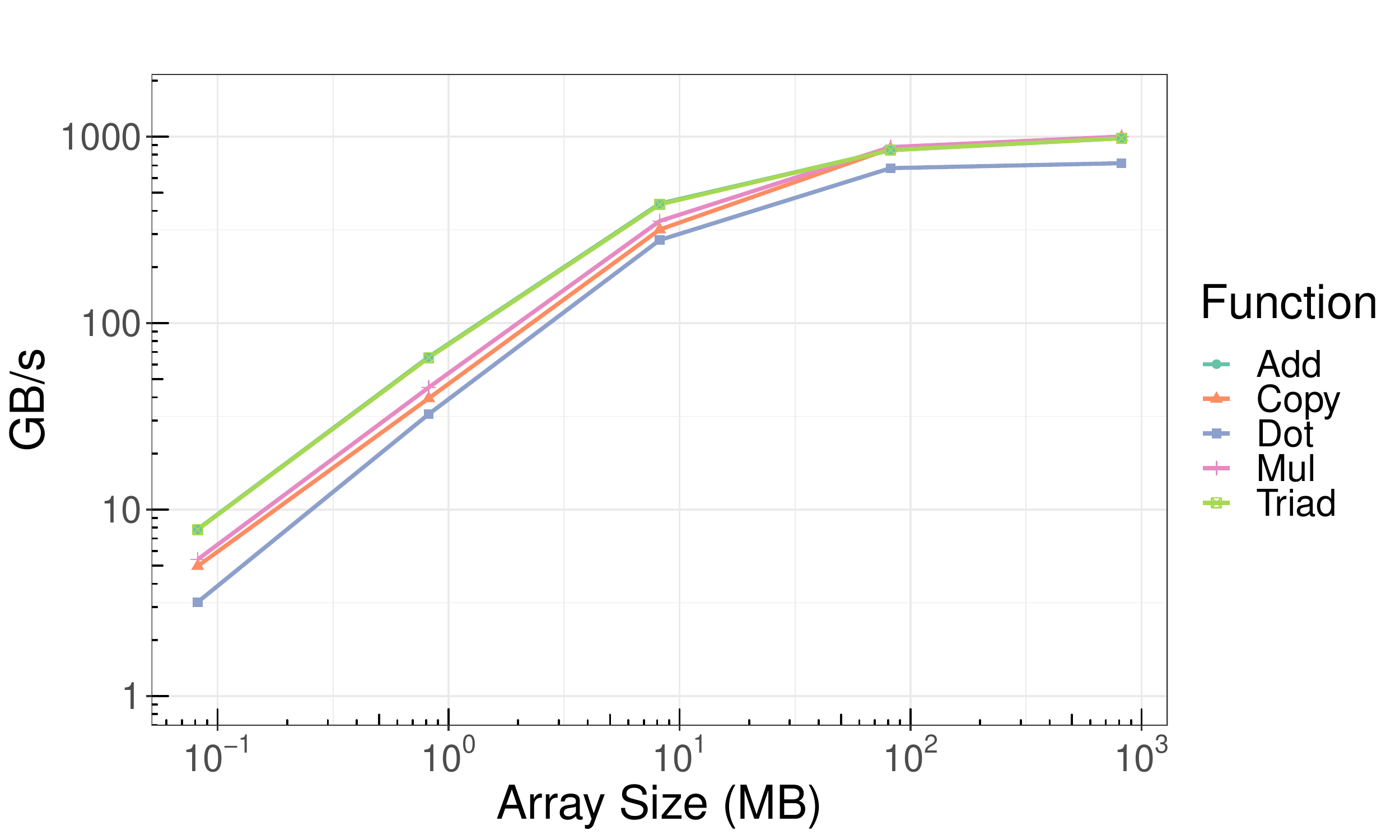}\\
    \includegraphics[width=.92\columnwidth]{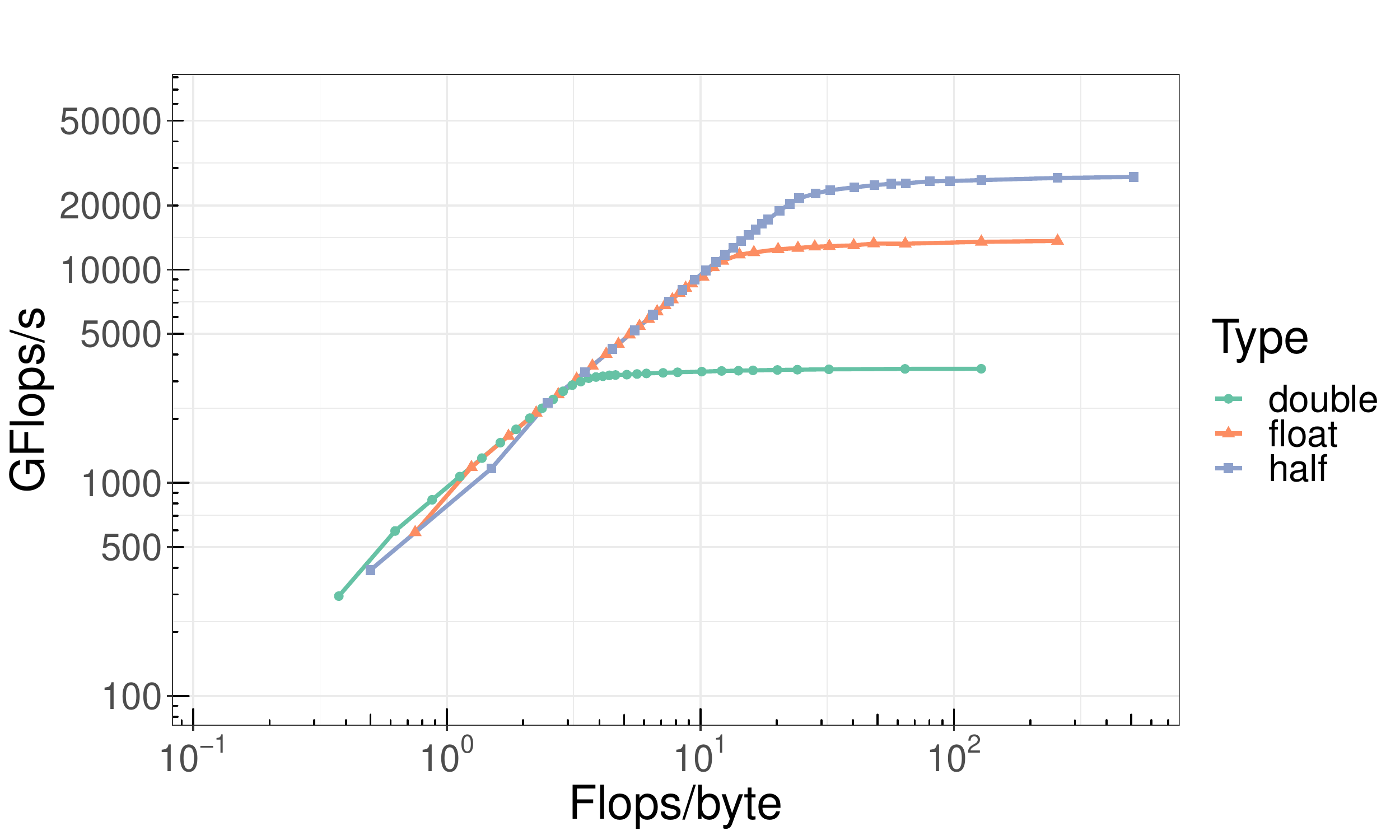}
    &
    \includegraphics[width=.92\columnwidth]{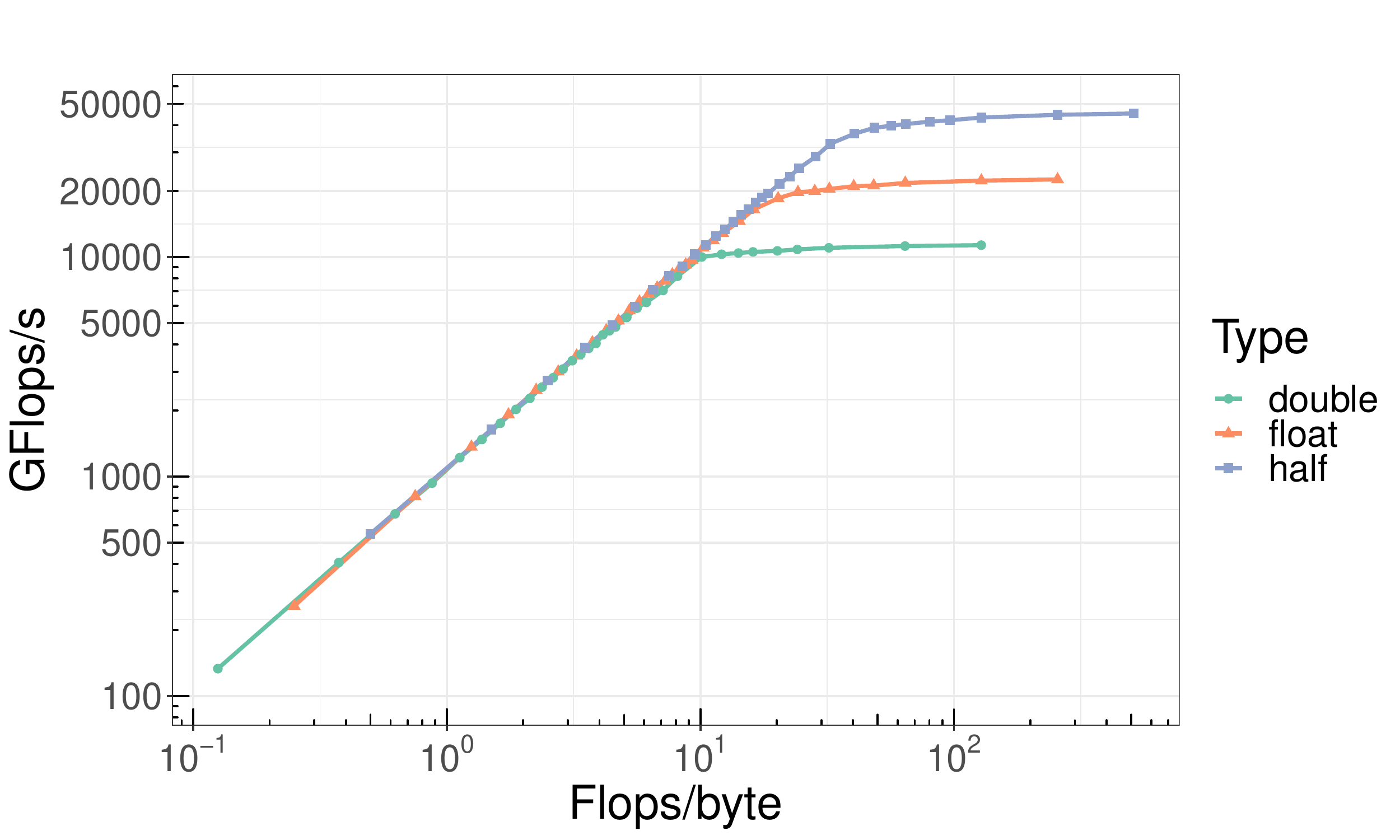}
  \end{tabular}
  \caption{Performance evaluation of the AMD RadeonVII GPU (left) and the AMD MI100 GPU (right) using the BabelStream benchmark (top) and the mixbench benchmark (bottom).}
  \label{fig:amdperf}
\end{figure*}

Overall, we advertise the following improvements that our custom conversion script provides compared to the plain {\sc DPCT} tool:
\begin{itemize}
\setlength\itemsep{.1em}
    \item As previously detailed, we generate an intermediate CUDA host function which invokes the kernel with the correct configurations in order to keep a code structure that is more similar to CUDA (see Figures~\ref{fig:dpctconversion}~and~\ref{fig:dpctconversion_improved});
    \item To generate a code which is similar to CUDA, we introduce a DPC++ kernel interface layer, that abstracts the differences to a CUDA or HIP kernel such as:
    \begin{itemize}
        \item We create a DPC++ \texttt{dim3} type to keep the CUDA kernel configuration and kernel call syntax consistent across the distinct executors;
        \item We implemented our own cooperative group environment as this functionality is currently not directly supported in DPC++;
        \item As DPC++ kernels need to access the kernel call configuration, we forward this information through the interface;
    \end{itemize}
    \item We automatically comment out in the original CUDA code any \texttt{sync()} functions as these would cause problems in the DPC++ conversion, which are then correctly replaced by our script;
    \item We disable some of the template instantiations (especially calling a macro which takes as parameter another macro and other arguments) as these would cause {\sc DPCT} errors.
    \item As \texttt{static\_cast<Type>(kernel-index)} in templated kernels fails to be converted correctly, we fix these expressions;
    \item We automatically convert the \gko CUDA-specific namespaces, variable names, types, \ldots to equivalent names containing DPC++ (e.g., in Figure~\ref{fig:dpctconversion_improved} there are still several mentions of CUDA remaining);
    \item We also automatically move the generated DPC++ kernel file into the correct \gko location.
\end{itemize}



\section{Performance Survey}
\label{sec:performance}

\begin{table}[!h]
    \centering
    \scriptsize
    \begin{tabular}{llccc}
    \hline
    \hline
     Name & Description & Programming Lang. & Release\\
     \hline
    NVIDIA V100 & Discrete server GPU & CUDA & 2017 \\
    NVIDIA A100 & Discrete server GPU & CUDA & 2020\\
    AMD RadeonVI & Discrete consumer GPU &  HIP & 2019 \\
    AMD MI100 & Discrete HPC GPU &  HIP & 2020 \\
    Intel Gen. 9 & Integrated GPU & DPC++ &  2015\\
    \hline
    \hline
    \end{tabular}
    \caption{List of the GPU architectures we consider in the performance evaluation.}
    \label{tab:architectures}
\end{table}

\begin{figure}
  \centering
    \includegraphics[width=.92\columnwidth]{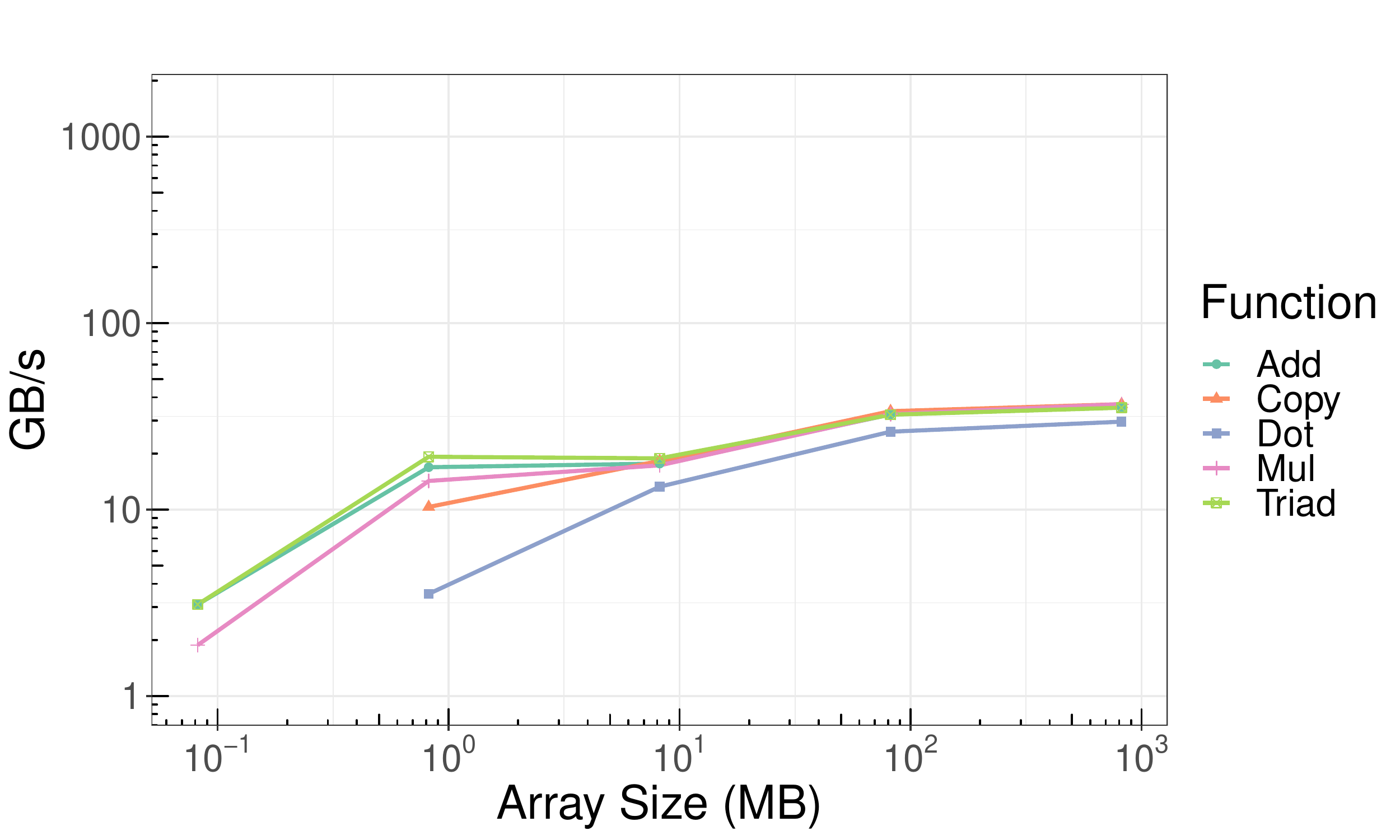}
  \caption{Performance evaluation of the AMD Intel Gen. 9 GPU using the BabelStream benchmark.}
  \label{fig:intelperf}
\end{figure}

The aspiration of \gko is to not only provide platform portability, but also a satisfying level of performance portability. A good indicator to assess whether this goal is achieved is to quantify the performance \gko achieves on different hardware architecture relative to the hardware-specific performance bounds. We acknowledge that the Reference executor is designed to check the correctness of the algorithms and provide a reference solution for the unit tests, and that \gko's primary focus is on high performance accelerators. Thus, we limit the performance analysis to the CUDA executor running on high-end NVIDIA GPUs, the HIP executor running on high-end AMD GPUs, and the DPC++ executor running on Intel integrated GPUs.
It is important to relate \gko's performance to the hardware-specific performance bounds. Thus, before providing performance results for \gko's routines, we present in \Cref{fig:nvidiaperf}, \Cref{fig:amdperf}, and \Cref{fig:intelperf} the results of the mixbench~\cite{mixbench} and BabelStream~\cite{babelstream} open source benchmarks.
Table~\ref{tab:architectures} lists the GPU hardware architectures we use in the performance evaluation along with a short architecture description. We note that not all architectures are high end server-line GPUs. The AMD RadeonVII is a consumer-line GPU, the Intel Gen. 9 GPU is an embedded GPU, however already supporting the DPC++ execution model. In the performance evaluation, we use CUDA v. 11.0 for the NVIDIA GPUs, ROCm v. 3.8 for the AMD GPUs, and DPC++ v. 2021.1-beta10 for the Intel GPUs.

The bandwidth analysis for the NVIDIA V100 ad NVIDIA A100 GPUs reveals a significant peak bandwidth improvement from the NVIDIA V100 GPU (peak bandwidth ~920 GB/s) to the NVIDIA A100 GPU (peak bandwidth ~1,400 GB/s), see top row in \Cref{fig:nvidiaperf}. At the same time, the bandwidth for small and moderate-sized data reads is higher on the V100 GPU. For the arithmetic performance, we note that the benchmark does not leverage the tensor cores on the V100 or the A100 GPU. For the general GPU cores, the peak performance of the A100 is for IEEE754 double precision almost exactly twice the V100 performance. For IEEE754 half precision, the performance improvements are smaller, and the IEEE754 single precision performance remains almost unchanged, see the bottom row in \Cref{fig:nvidiaperf}.

The two AMD GPU architectures we consider are comparable in their measured bandwidth: 850 GB/s for the AMD RadeonVII GPU vs 1,000 GB/s for the AMD MI100 GPU, see top row in \Cref{fig:amdperf}. At the same time, they significantly differ in their arithmetic performance, see bottom row in \Cref{fig:amdperf}.

At the time of writing the mixbench benchmark does not provide a DPC++ version for the execution on Intel GPUs. Thus, in \Cref{fig:intelperf}, we only report the bandwidth performance of the considered Intel Gen. 9 GPU. We note that the performance for the dot kernel is about 10\% below the maximum achievable bandwidth. For the other operations, the bandwidth plateaus for array sizes exceeding 100 MB at a rate of 35 GB/s.

\begin{figure*}[!h]
  \centering
  \begin{tabular}{lr}
    \includegraphics[width=.92\columnwidth]{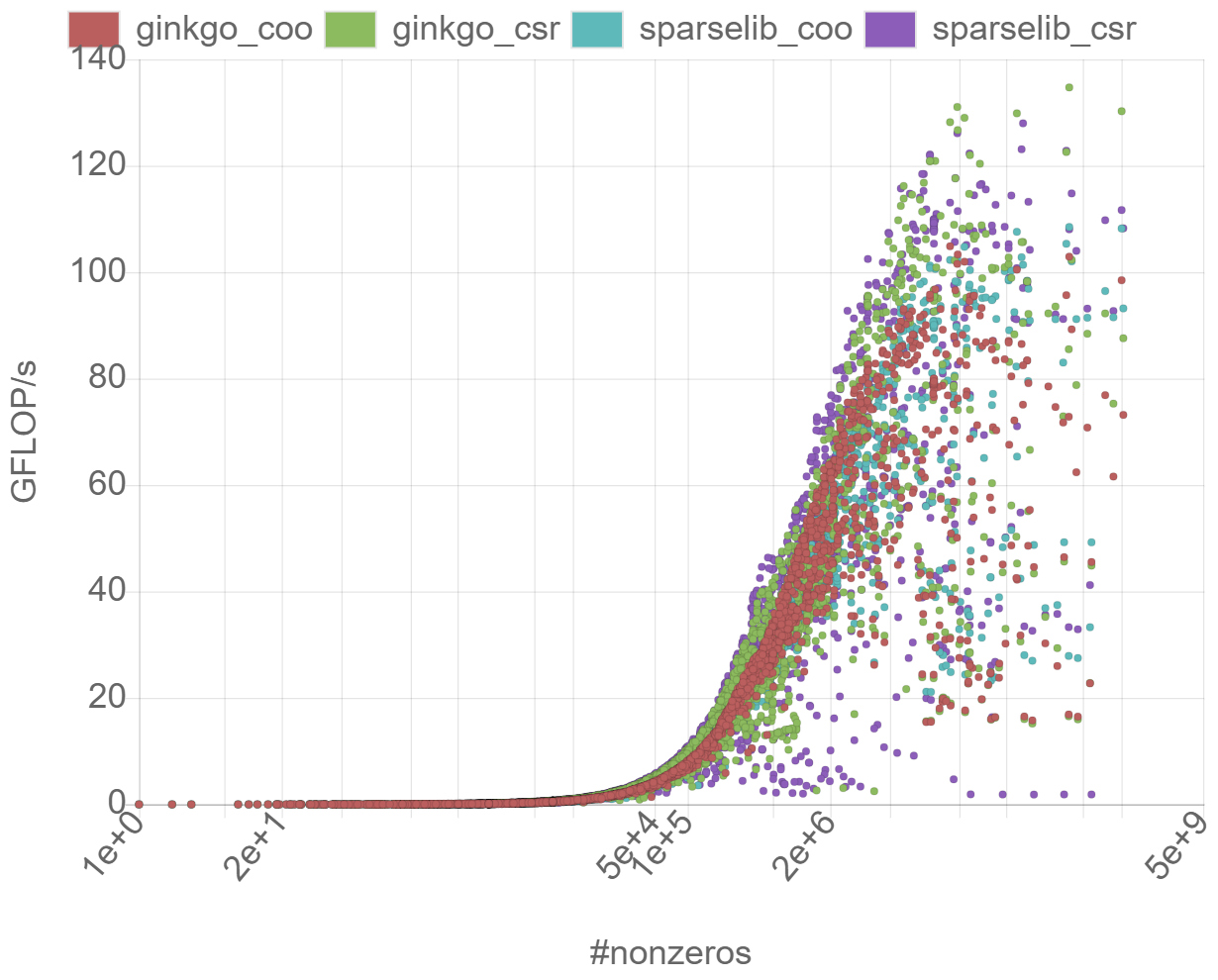}
    &
    \includegraphics[width=.92\columnwidth]{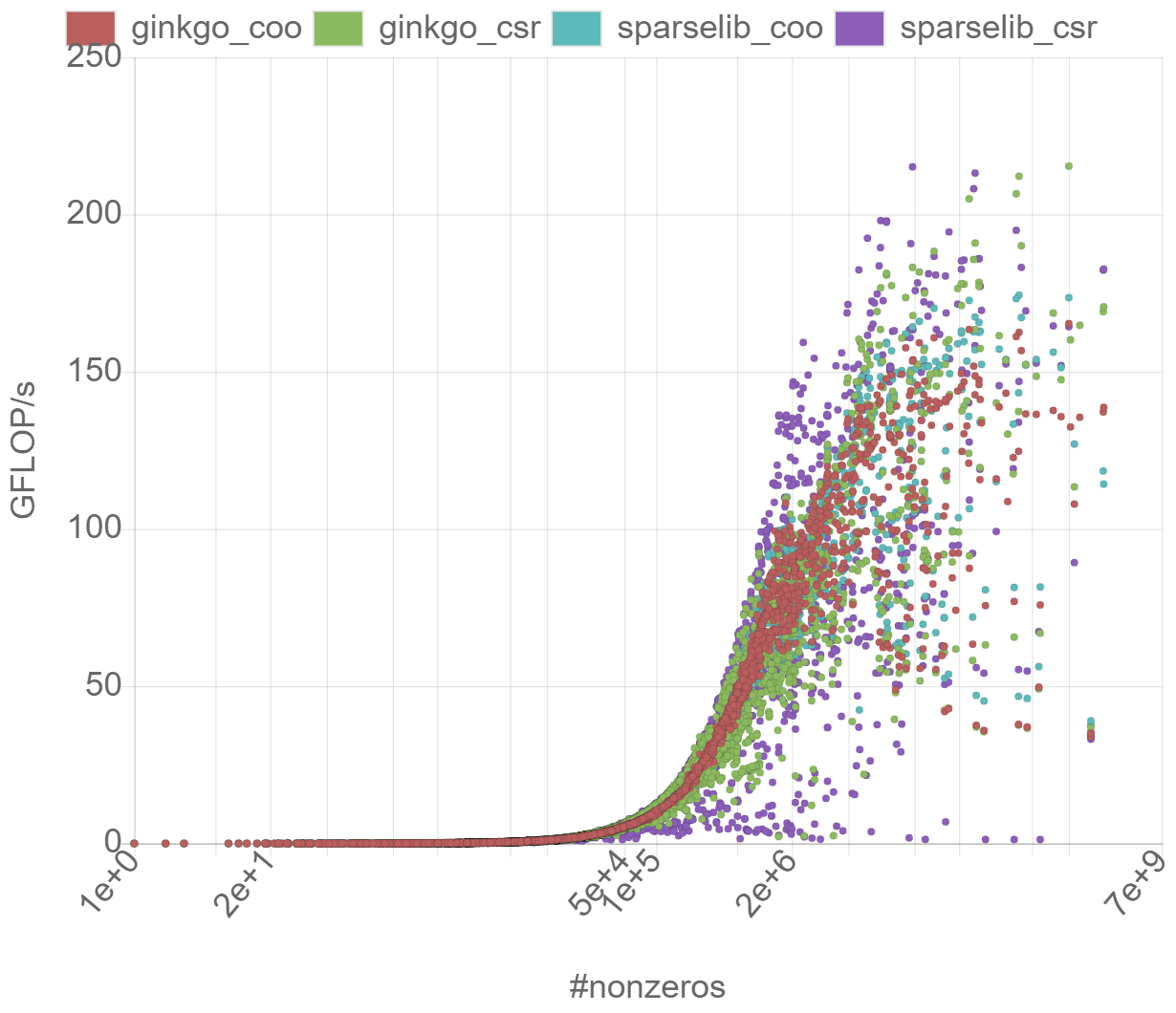}
  \end{tabular}
  \caption{Performance of the \gko SpMV and the cuSPARSE SpMV (sparselib) on the NVIDIA V100 GPU (left) and the NVIDIA A100 GPU (right).}
  \label{fig:cudaspmv}
\end{figure*}

\begin{figure*}[!h]
  \centering
  \begin{tabular}{lr}
    \includegraphics[width=.92\columnwidth]{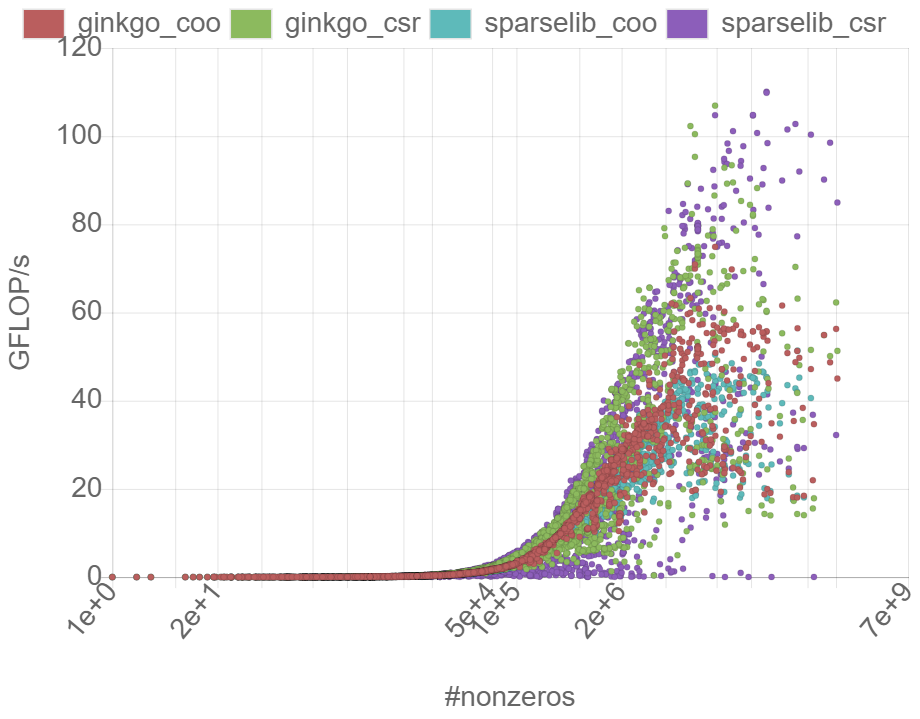}
    &
    \includegraphics[width=.92\columnwidth]{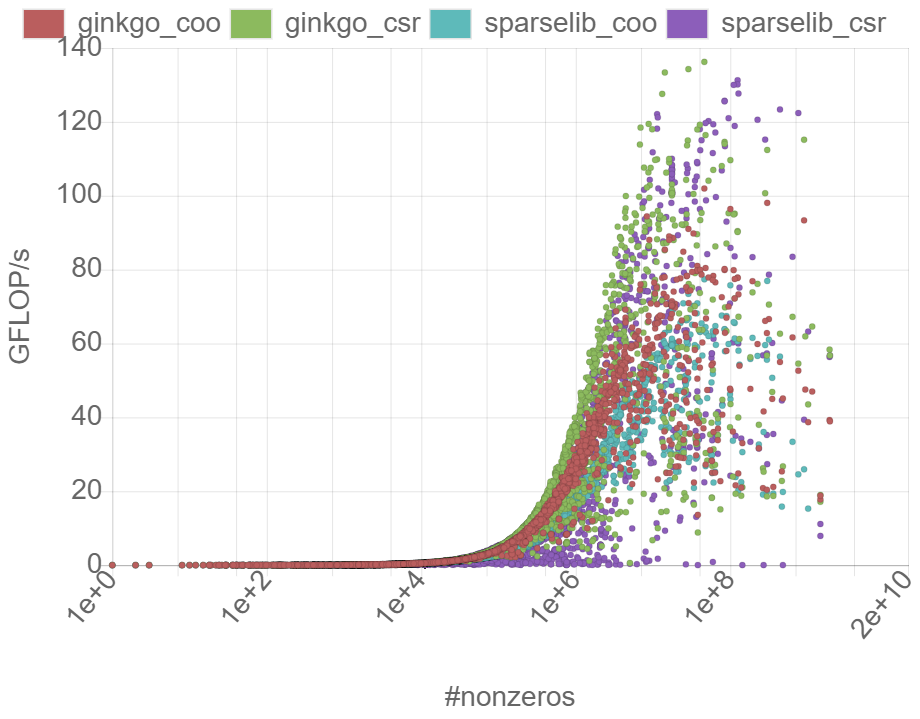}
  \end{tabular}
  \caption{Performance of the \gko SpMV and the hipSPARSE SpMV (sparselib) on the AMD RadeonVII GPU (left) and the AMD MI100 GPU (right).}
  \label{fig:hipspmv}
\end{figure*}

\subsection{Ginkgo SpMV performance}
We first investigate the performance \gko's sparse matrix vector product kernels (SpMV) kernels achieve on the distinct executors. For this, we take 100 representative matrices from the Suite Sparse Matrix Collection~\cite{suitesparse} as benchmark, and run a set of heavily-tuned sparse matrix vector product kernels on the distinct architectures. All computations use double precision arithmetic. The distinct SpMV kernels differ in terms of how they store the sparse matrix and which processing strategy they apply~\cite{spmvtopc}. In general, we may expect a performance peak of (memory bandwidth) / (8byte/entry + 4 byte/entry + 4 byte/entry) * 2 ops/entry = (memory bandwidth) / 8 for the COO SpMV kernels, and (memory bandwidth) / 6 for the CSR kernel.

In \Cref{fig:cudaspmv} we report the performance of different SpMV kernels taken from either \gko or the vendor library (NVIDIA cuSPARSE~\cite{cuda}) on the NVIDIA V100 GPU (left) and NVIDIA A100 GPU (right). Each dot represents one combination of SpMV kernel and test matrix. We cannot identify a clear winner in the performance graphs, which is expected as the distinct kernels differ in their efficiency for the distinct problem characteristics. However, we can identify the \gko kernels to be competitive to the cuSPARSE SpMV kernels. The maximum performance numbers achieved are around 135 GFLOP/s and 220 GFLOP/s on the NVIDIA V100 and NVIDIA A100 GPUs. We notice that this performance difference exceeds the expectations based on the bandwidth improvements of 1.5$\times$. At the same time, the performance is very close to the expected upper bound of approximately 150 GFLOP/s and 230 GFLOP/s, respectively.

\begin{figure}[!h]
  \centering
    \includegraphics[width=.92\columnwidth]{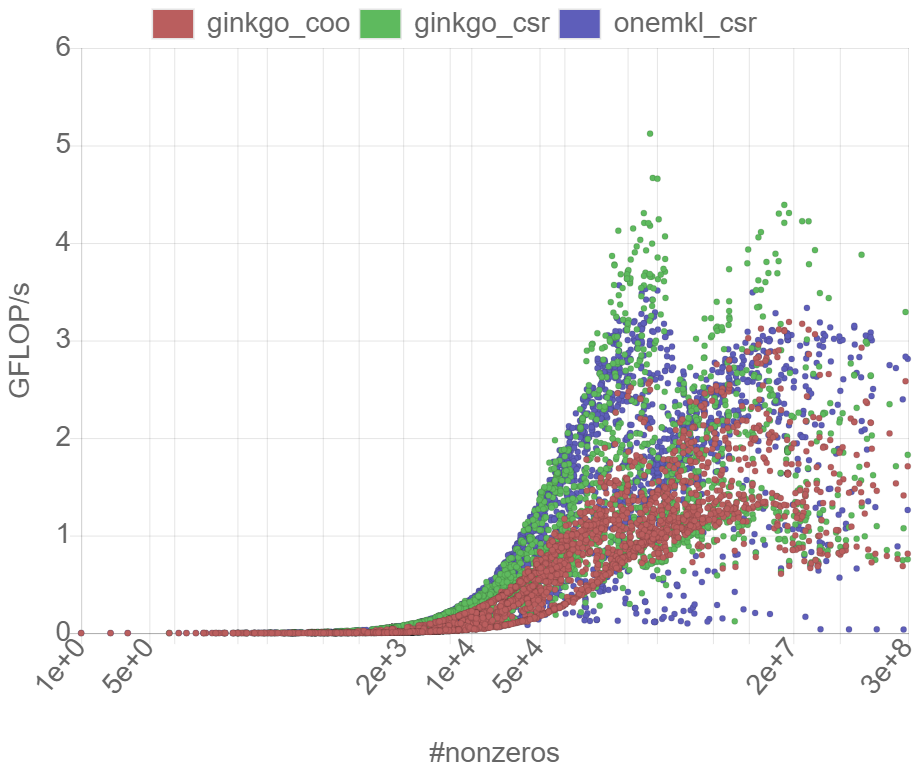}
  \caption{Performance of the \gko SpMV and the intel MKL SpMV on the Intel Gen. 9 GPU.}
  \label{fig:intelspmv}
\end{figure}

In \Cref{fig:hipspmv}, we present performance results obtained from running SpMV benchmarks on AMD hardware, the AMD RadeonVII GPU (left) and the AMD MI100 GPU (right). We use the same experimental setup and again include both SpMV routines from the vendor library (hipSPARSE~\cite{hip}) and from \gko. For both architectures, there is no clear performance winner, but we recognize \gko's SpMV kernels being highly competitive to the vendor library hipSPARSE. We observe the \gko and sparselib CSR SpMV kernels achieve the highest performance numbers, up to 110 GFLOP/s on the AMD RadeonVII GPU and up to 138 GFLOP/s on the AMD MI100 GPU. This is about 80\% of the bandwidth-induced theoretical performance bound. We advertise that the rocSPARSE library may provide higher performance than hipSPARSE, but we report the performance of the latter since our library focuses on the portable HIP layer. Furthermore, we acknowledge that that the hardware and software stack for the AMD MI100 GPU is still under development, and further performance increase can be expected.

\Cref{fig:intelspmv} presents the performance results we obtain from running the \gko SpMV on the Intel Gen. 9 GPU. As expected from the significantly lower bandwidth, the SpMV performance is only a fraction of the performance achieved on AMD and NVIDIA performance, which is expected given the embedded design of the Gen. 9 GPU. However, with a peak of 5 GFLOP/s for the \gko CSR SpMV kernel  the performance is extremely close to the bandwidth-induced theoretical peak. We also note the \gko SpMV generally outperforming the Intel MKL CSR SpMV

\subsection{Ginkgo solver performance}
Next, we assess the performance of \gko's Krylov subspace solvers achieve on the distinct executors on recent hardware architectures from AMD, NVIDIA, and Intel. For the performance evaluation, we select 10 test matrices that are different in size and origin to cover a wide spectrum of applications. We note that all Krylov solvers we consider are memory bound, and hence the performance being limited by the memory bandwidth. If we ignore the column and row indexing information in the sparse matrix vector product and use the aggressive approximation of an arithmetic intensity $ai = 1$, the performance of the algorithms is the ratio between memory bandwidth (in GB/s) and the precision format complexity (8 byte per floating point value for IEEE754 double precision). All solvers are run for 10,000 iterations to account for machine noise. The \gko solvers employ the \gko COO SpMV for generating the Krylov subspace.

\begin{figure*}[!h]
  \centering
  \begin{tabular}{lr}
    \includegraphics[width=.94\columnwidth]{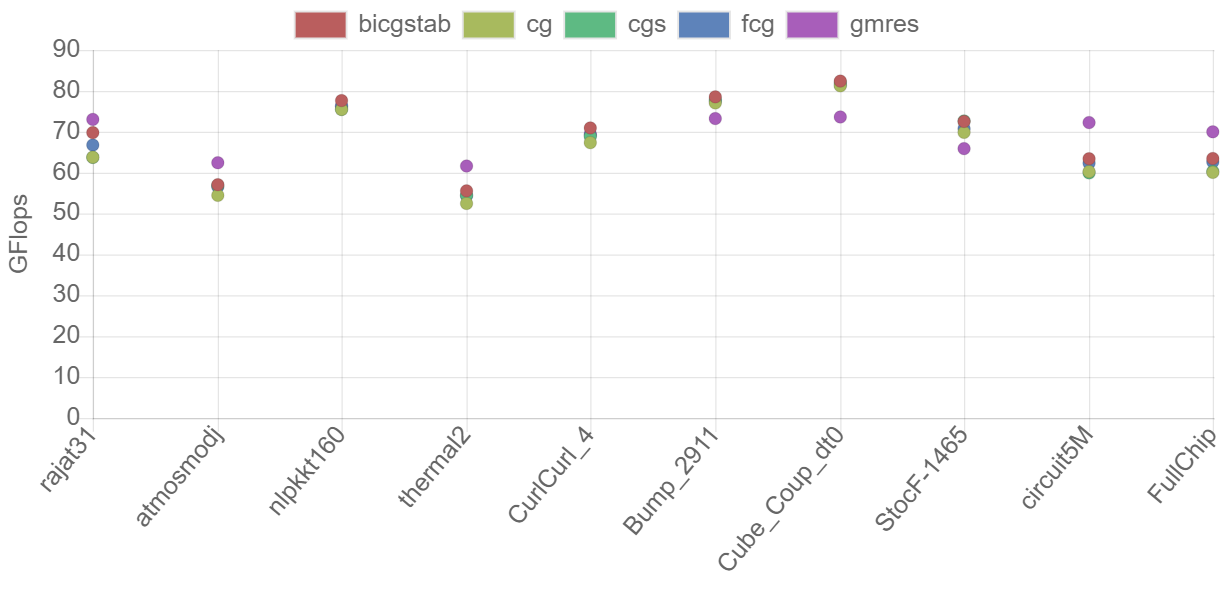}
    &
    \includegraphics[width=.94\columnwidth]{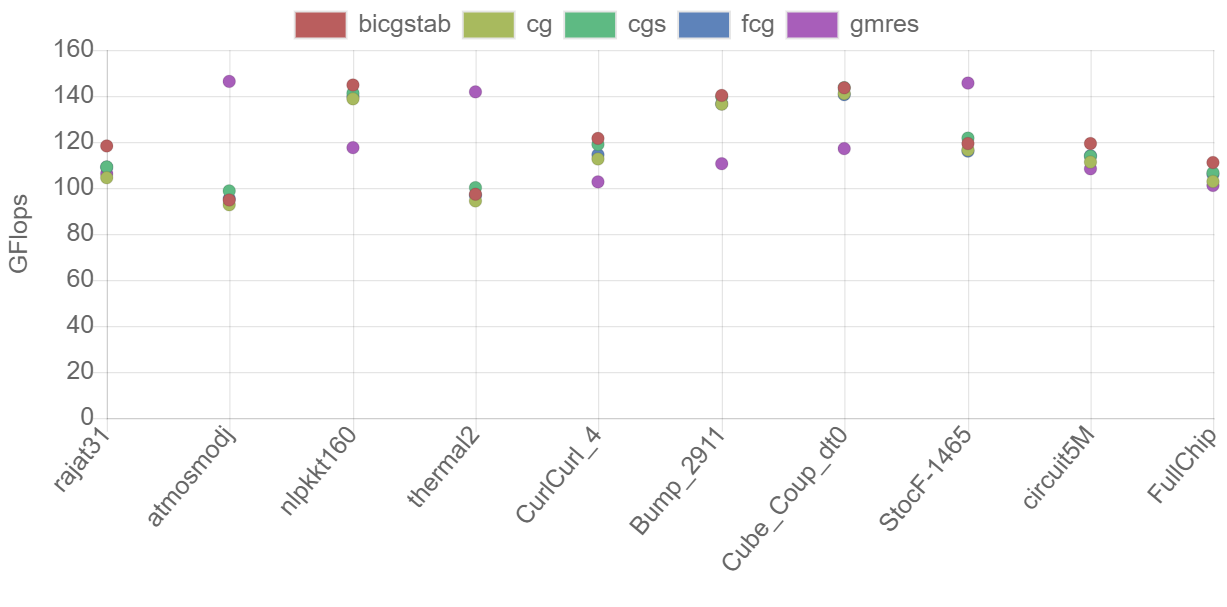}
  \end{tabular}
  \caption{Krylov solver performance of the \gko CUDA backend on the NVIDIA V100 GPU (left) and the NVIDIA A100 GPU (right).}
  \label{fig:cudasolver}
\end{figure*}

As the CUDA executor was developed as \gko's first high performance GPU backend, we report in \Cref{fig:cudasolver} the performance of a subset of \gko's Krylov solvers on the NVIDIA V100 GPU (left) and on the newer NVIDIA A100 GPU (right). On the over V100 architecture, the performance of the Krylov solvers varies between 50 GFLOP/s and 80 GFLOP/s, depending on the specific problem and solver combination. The median of 65 GFLOP/s is significantly lower than the median of 100 GFLOP/s on the newer A100 GPU, see the right-hand side of \Cref{fig:cudasolver}. On the A100 GPU, the performance of all solvers exceeds 80 GFLOP/s, independent of the specific test matrix, and achieve up to 140 GFLOP/s for some of the test problems. This is extremely close (82\%) to the very aggressive theoretical upper bound approximation of 170 GFLOP/s.

\begin{figure*}[!h]
  \centering
  \begin{tabular}{lr}
    \includegraphics[width=.94\columnwidth]{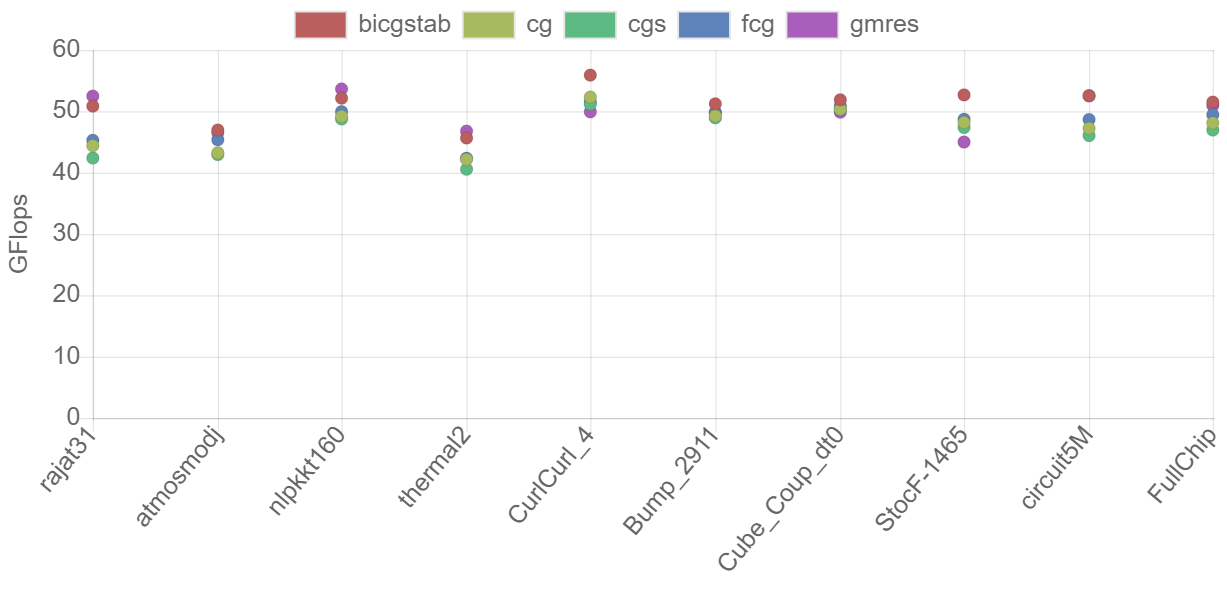}
    &
    \includegraphics[width=.94\columnwidth]{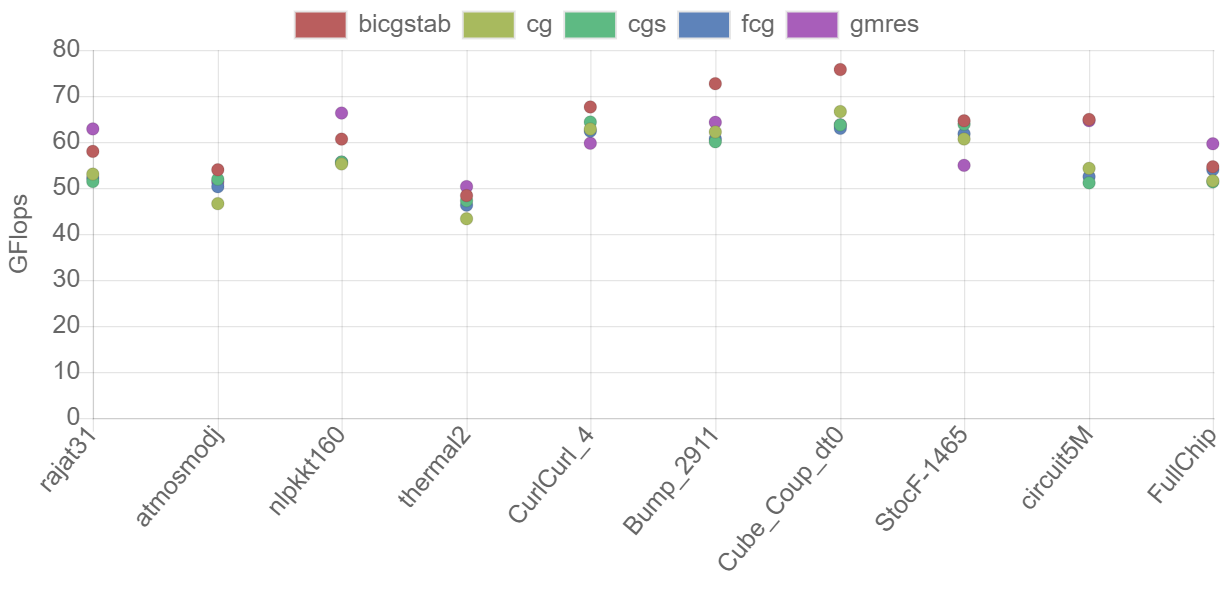}
  \end{tabular}
  \caption{Krylov solver performance of the \gko HIP backend on the AMD RadeonVII GPU (left) and the AMD MI100 GPU (right).}
  \label{fig:hipsolver}
\end{figure*}

Even though originally focused on NVIDIA GPU execution via the CUDA executor, we can from the similarity in architecture design expect the algorithms to achieve high efficiency values also on the AMD architectures. In \Cref{fig:hipsolver} we report the \gko Krylov solver performance on the AMD Radeon VII GPU (left) and the AMD MI100 GPU (right). These results are obtained using the HIP executor designed for AMD backends. Aside from the hardware-specific parameter configurations, the HIP kernels invoked by the Krylov solvers are very similar to the CUDA kernels used by the CUDA exector. The evaluation on the AMD RadeonVII GPU reveals the same problem-specific performance variation that we observed in the CUDA backend evaluation. The median performance of the Krylov solvers is around 50 GFLOP/s with performance variations between 40 GFLOP/s and 55 GFLOP/s. On the newer server-line AMD MI100 GPU, the performance increases to a range between 42 GFLOP/s and 77 GFLOP/s, with a median around 65 GFLOP/s. Using the same aggressive approximation for the arithmetic intensity, the theoretical performance bounds approximate 100 GFLOP/s for the AMD RadeonVII and 120 GFLOP/s for the AMD MI100, respectively.
We acknowledge that the HIP executor achieves a smaller fraction of the theoretical performance bound, however, with achieving $>$50\% of the theoretical peak, we still claim performance portability of the algorithms.

\begin{figure}[!h]
  \centering
    \includegraphics[width=.94\columnwidth]{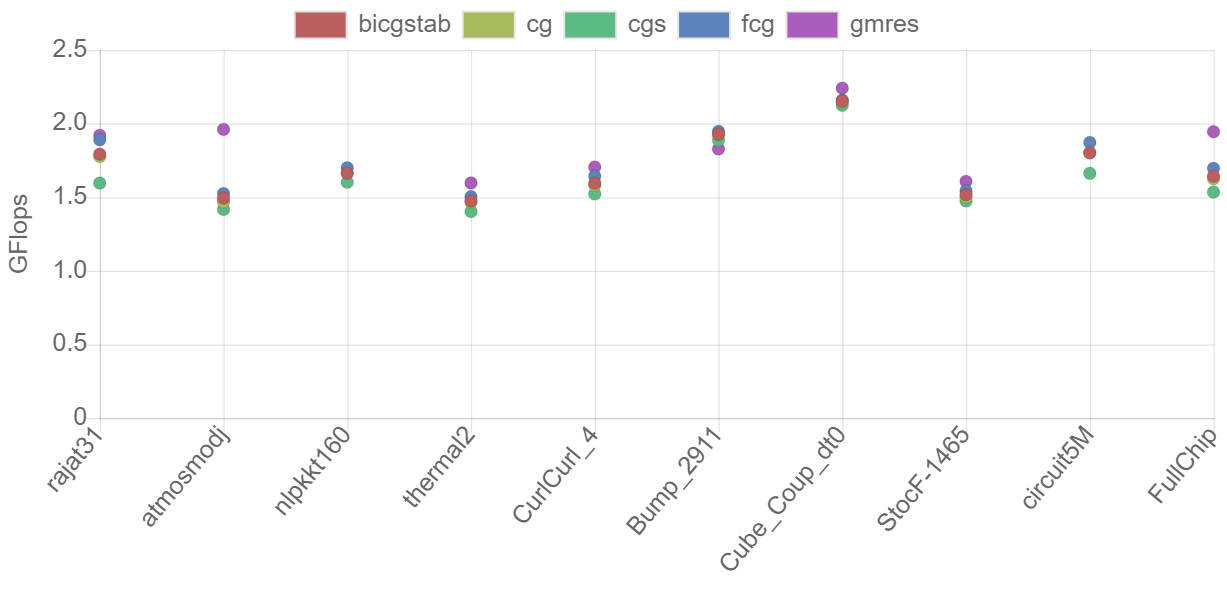}
  \caption{Krylov solver performance of the \gko DPC++ backend on the Intel Gen. 9 GPU.}
  \label{fig:intelsolver}
\end{figure}

From the Intel GPU bandwidth analysis, we expect \gko{}'s Krylov solvers to achieve up to 4.2 GFLOP/s. The results in\Cref{fig:intelsolver} reveal the actual performance results being significantly lower, only achieving 1.4 -- 2 GFLOP/s on the Intel Gen. 9 GPU. This may be attributed to \gko{}'s DPC++ executor being in his early stages, and the OneAPI ecosystem still being under development. We, again, note that the Intel Gen. 9 GPU is an integrated GPU, and not expected to achieve high performance. The purpose of this exercise is rather to show the validity of \gko{}'s design, and the technology readiness for the Intel high end GPU platform to come.

\section{Summary and Outlook}
\label{sec:conclusion}
In this paper, we elaborate on how \gko tackles platform portability by separating the numerical core from the hardware-specific backends. We discuss how we adopt the execution space to NVIDIA GPUs via the CUDA language, AMD GPUs via the HIP language, and Intel GPUs via the DPC++ language. We also report performance results for running basic sparse linear algebra operations and complete Krylov solvers on the newest hardware architectures from these vendors, and demonstrate \gko's performance portability and identify \gko's sparse matrix vector product being highly competitive or even outperforming the vendor libraries.

\bigskip

\textbf{Acknowledgments}
This work was supported by the US Exascale Computing Project
(17-SC-20-SC), a collaborative effort of the U.S. Department of Energy Office
of Science and the National Nuclear Security Administration.

\bibliography{references}

\end{document}